\documentclass[12pt]{article}
\title{Randomized selection with tripartitioning}
\author{Krzysztof C. Kiwiel\thanks{Systems Research Institute,
        Newelska 6, 01--447 Warsaw, Poland
        ({\tt kiwiel@ibspan.waw.pl})}}
\date{January 4, 2004}

\newcommand{\BbbF}{{\rm\normalcolor I\kern-.18em F}}
\newcommand{\BbbR}{{\rm\normalcolor I\kern-.18em R}}
\newcommand{\eqref}[1]{{\normalfont\normalcolor(\ref{#1})}}
\makeatletter
\def\proof{%
   \def\a##1{\begin{trivlist}\item[]{\bf\ignorespaces{##1}.}%
    \enspace\ignorespaces}%
   \def\b[##1]{\a{Proof\ \ignorespaces{##1}}}%
   \@ifnextchar[{\b}{\a{Proof}}}
\def\endproof{\end{trivlist}}
\def\qed{\relax\protect\ifmmode\ifinner\else\quad\fi\fi
    \hbox{\vbox{\hrule height.4pt\hbox{\vbox{\hrule height.4pt
    \hbox{\vrule width.4pt\vphantom{\normalsize A}\kern.5em
    \vrule width.4pt}\hrule height.4pt}}}}}
\newtoks\@stequation
\def\subequations{\refstepcounter{equation}%
\edef\@savedequation{\the\c@equation}%
\@stequation=\expandafter{\theequation}%   %only want \theequation
\edef\@savedtheequation{\the\@stequation}% %expanded once
\edef\oldtheequation{\theequation}%
\setcounter{equation}{0}%
\def\theequation{\oldtheequation\alph{equation}}}%
\def\endsubequations{%
\setcounter{equation}{\@savedequation}%
\@stequation=\expandafter{\@savedtheequation}%
\edef\theequation{\the\@stequation}\global\@ignoretrue}
\def\@begintheorem#1#2{\trivlist
    \item[\hskip \labelsep{\bfseries #1\ #2.}]\itshape}
\def\@opargbegintheorem#1#2#3{\trivlist
    \item[\hskip \labelsep{\bfseries #1\ #2\ (#3).}]\itshape}
\@addtoreset{equation}{section}% Makes \section reset `equation' counter.
\def\theequation{\thesection.\arabic{equation}}
\@addtoreset{figure}{section}

\@addtoreset{table}{section}

\let\@@eqnsel=\relax
\def\@tempa{%
    \stepcounter{equation}%
    \def\@currentlabel{\p@equation\theequation}%
    \global\@eqnswtrue\m@th
    \global\@eqcnt\z@
    \tabskip\mathindent
    \let\\=\@eqncr
    \setlength\abovedisplayskip{\topsep}%
    \ifvmode
      \addtolength\abovedisplayskip{\partopsep}%
    \fi
    \addtolength\abovedisplayskip{\parskip}%
    \setlength\belowdisplayskip{\abovedisplayskip}%
    \setlength\belowdisplayshortskip{\abovedisplayskip}%
    \setlength\abovedisplayshortskip{\abovedisplayskip}%
    $$\everycr{}\halign to\linewidth% $$
    \bgroup
      \hskip\@centering
      $\displaystyle\tabskip\z@skip{##}$\@eqnsel&%
      \global\@eqcnt\@ne \hskip \tw@\arraycolsep \hfil${##}$\hfil&%
      \global\@eqcnt\tw@ \hskip \tw@\arraycolsep
        $\displaystyle{##}$\hfil \tabskip\@centering&%
      \global\@eqcnt\thr@@
        \hb@xt@\z@\bgroup\hss##\egroup\tabskip\z@skip\cr}%
\def\@tempb{%
   \stepcounter{equation}%
   \def\@currentlabel{\p@equation\theequation}%
   \global\@eqnswtrue
   \m@th
   \global\@eqcnt\z@
   \tabskip\@centering
   \let\\\@eqncr
   $$\everycr{}\halign to\displaywidth\bgroup
       \hskip\@centering$\displaystyle\tabskip\z@skip{##}$\@eqnsel
      &\global\@eqcnt\@ne\hskip \tw@\arraycolsep \hfil${##}$\hfil
      &\global\@eqcnt\tw@ \hskip \tw@\arraycolsep
         $\displaystyle{##}$\hfil\tabskip\@centering
      &\global\@eqcnt\thr@@ \hb@xt@\z@\bgroup\hss##\egroup
         \tabskip\z@skip
      \cr
}
\ifx\eqnarray\@tempa%     If the fleqn document-class option is in effect
    \def\eqnarray{%
    \stepcounter{equation}%
    \def\@currentlabel{\p@equation\theequation}%
    \global\@eqnswtrue\m@th
    \global\@eqcnt\z@
    \tabskip\mathindent
    \let\\=\@eqncr
    \setlength\abovedisplayskip{\topsep}%
    \ifvmode
      \addtolength\abovedisplayskip{\partopsep}%
    \fi
    \addtolength\abovedisplayskip{\parskip}%
    \setlength\belowdisplayskip{\abovedisplayskip}%
    \setlength\belowdisplayshortskip{\abovedisplayskip}%
    \setlength\abovedisplayshortskip{\abovedisplayskip}%
    $$\everycr{}\halign to\linewidth% $$
    \bgroup
      \hskip\@centering
      $\displaystyle\tabskip\z@skip{##}$\@eqnsel&%
      \global\@eqcnt\@ne
      \@@eqnsel%            \@@eqnsel has replaced \hskip \tw@\arraycolsep!!!
      \hfil${{}##{}}$\hfil&%              as in fixup.sty but textstyle!!!
      \global\@eqcnt\tw@
      \@@eqnsel%           \@@eqnsel has replaced \hskip \tw@\arraycolsep!!!
        $\displaystyle{##}$\hfil \tabskip\@centering&%
      \global\@eqcnt\thr@@
        \hb@xt@\z@\bgroup\hss##\egroup\tabskip\z@skip\cr}%
\else\ifx\eqnarray\@tempb%       Else try the default eqnarray environment.
   \def\eqnarray{%
   \stepcounter{equation}%
   \def\@currentlabel{\p@equation\theequation}%
   \global\@eqnswtrue
   \m@th
   \global\@eqcnt\z@
   \tabskip\@centering
   \let\\\@eqncr
   $$\everycr{}\halign to\displaywidth\bgroup
       \hskip\@centering$\displaystyle\tabskip\z@skip{##}$\@eqnsel
      &\global\@eqcnt\@ne
      \@@eqnsel%           \@@eqnsel has replaced \hskip \tw@\arraycolsep!!!
      \hfil${{}##{}}$\hfil%              as in fixup.sty but textstyle!!!
      &\global\@eqcnt\tw@
      \@@eqnsel%           \@@eqnsel has replaced \hskip \tw@\arraycolsep!!!
         $\displaystyle{##}$\hfil\tabskip\@centering
      &\global\@eqcnt\thr@@ \hb@xt@\z@\bgroup\hss##\egroup
         \tabskip\z@skip
      \cr}
\else 
\fi\fi
\def\@tempa{}			% Free up TeX's memory
\def\@tempb{}
\@ifundefined{chapter}{%
  \renewenvironment{thebibliography}[1]
     {\section*{\refname
        \@mkboth{\MakeUppercase\refname}{\MakeUppercase\refname}}%
      \list{\@biblabel{\@arabic\c@enumiv}}%
           {\settowidth\labelwidth{\@biblabel{#1}}%
            \leftmargin\labelwidth
            \advance\leftmargin\labelsep
            \itemsep \z@                 % Suppresses vertical separation.
            \@openbib@code
            \usecounter{enumiv}%
            \let\p@enumiv\@empty
            \renewcommand\theenumiv{\@arabic\c@enumiv}}%
      \sloppy
      \clubpenalty4000
      \@clubpenalty \clubpenalty
      \widowpenalty4000%
      \sfcode`\.\@m}
     {\def\@noitemerr
       {\@latex@warning{Empty `thebibliography' environment}}%
      \endlist}}%
{\renewenvironment{thebibliography}[1]
     {\section*{\bibname
        \@mkboth{\MakeUppercase\bibname}{\MakeUppercase\bibname}}%
      \list{\@biblabel{\@arabic\c@enumiv}}%
           {\settowidth\labelwidth{\@biblabel{#1}}%
            \leftmargin\labelwidth
            \advance\leftmargin\labelsep
            \itemsep \z@                 % Suppresses vertical separation.
            \@openbib@code
            \usecounter{enumiv}%
            \let\p@enumiv\@empty
            \renewcommand\theenumiv{\@arabic\c@enumiv}}%
      \sloppy
      \clubpenalty4000
      \@clubpenalty \clubpenalty
      \widowpenalty4000%
      \sfcode`\.\@m}
     {\def\@noitemerr
       {\@latex@warning{Empty `thebibliography' environment}}%
      \endlist}}%
\newcommand{\Argmax}{{\operator@font Arg}\max}
\newcommand{\Argmin}{{\operator@font Arg}\min}
\newcommand{\argmax}{{\operator@font arg}\max}
\newcommand{\argmin}{{\operator@font arg}\min}
\newcommand{\Exp}{\mathord{\operator@font E}}
\newcommand{\med}{\mathop{\operator@font med}}
\newcommand{\Prob}{\mathord{\operator@font P}}
\newcommand{\rank}{\mathop{\operator@font rank}}
\newcommand{\var}{\mathop{\operator@font var}}
\makeatother
\newtheorem{theorem}{Theorem}[section]
\newtheorem{algorithm}[theorem]{Algorithm}

\newtheorem{corollary}[theorem]{Corollary}

\newtheorem{fact}[theorem]{Fact}
\newtheorem{lemma}[theorem]{Lemma}

\newtheorem{remarks}[theorem]{Remarks}
\newtheorem{scheme}{Scheme}

\begin{document}           % End of preamble and beginning of text.

\maketitle                 % Produces the title.

\begin{abstract}
\noindent
We show that several versions of Floyd and Rivest's algorithm
{\sc Select} [Comm.\ ACM {\bf 18} (1975) 173] for finding the $k$th
smallest of $n$ elements require at most $n+\min\{k,n-k\}+o(n)$
comparisons on average, even when equal elements occur.  This parallels
our recent analysis of another variant due to Floyd and Rivest
[Comm.\ ACM {\bf 18} (1975) 165--172].  Our computational results
suggest that both variants perform well in practice, and may compete
with other selection methods, such as Hoare's {\sc Find} or
quickselect with median-of-3 pivots.
\end{abstract}

\begin{quotation}
\end{quotation}

%\begin{quotation}
%\noindent{\bf MSC Subject Classifications.} 68W20, 68W05, 68Q25
%\end{quotation}

%\begin{quotation}
%\noindent{\bf Abbreviated title:} Randomized selection.
%\end{quotation}

%   *** SECTION 1 ***
\section{Introduction}
\label{s:intro}
The {\em selection problem\/} is defined as follows: Given a set
$X:=\{x_j\}_{j=1}^n$ of $n$ elements, a total order $<$ on $X$,
and an integer $1\le k\le n$, find the {\em $k$th smallest\/}
element of $X$, i.e., an element $x$ of $X$ for which there are at
most $k-1$ elements $x_j<x$ and at least $k$ elements $x_j\le x$.
The {\em median\/} of $X$ is the $\lceil n/2\rceil$th smallest
element of $X$.

Selection is one of the fundamental problems in computer science;
see, e.g., the references in \cite{dohaulzw:lbs,dozw:sm,dozw:msr} and
\cite[\S5.3.3]{knu:acpIII2}.  Most references concentrate on the
number of comparisons between pairs of elements made in selection
algorithms.  In the worst case, selection needs at least
$(2+\epsilon)n$ comparisons \cite{dozw:msr}, whereas the algorithm of
\cite{blflprrita:tbs} makes at most $5.43n$, that of \cite{scpapi:fm}
needs $3n+o(n)$, and that in \cite{dozw:sm} takes $2.95n+o(n)$.  In the
average case, for $k\le\lceil n/2\rceil$, at least $n+k-O(1)$
comparisons are necessary \cite{cumu:acs}, whereas the best upper bound
is $n+k+O(n^{1/2}\ln^{1/2}n)$ \cite[Eq.\ (5.3.3.16)]{knu:acpIII2}.  The
classical algorithm {\sc Find} of \cite{hoa:a65}, also known as
quickselect, has an upper bound of $3.39n+o(n)$ for $k=\lceil n/2\rceil$
in the average case \cite[Ex.\ 5.2.2--32]{knu:acpIII2}, which improves
to $2.75n+o(n)$ for median-of-3 pivots \cite{gru:mvh,kimapr:ahf}.

In practice {\sc Find} is most popular.  One reason is that the
algorithms of \cite{blflprrita:tbs,scpapi:fm} are much slower on the
average \cite{mus:iss,val:iss}, whereas \cite{kimapr:ahf} adds that
other methods proposed so far, although better than {\sc Find} in
theory, are not practical because they are difficult to implement,
their constant factors and hidden lower order terms are too large,
etc.  It is quite suprising that these references
\cite{kimapr:ahf,mus:iss,val:iss} ignore the algorithm {\sc Select}
of \cite{flri:etb}, since most textbooks mention that {\sc Select} is
asymptotically faster than {\sc Find}.  In contrast, this paper shows
that {\sc Select} can compete with {\sc Find} in both theory and
practice, even for fairly small values of the input size $n$.

We now outline our contributions in more detail.  The initial two
versions of {\sc Select} \cite{flri:etb} had gaps in their analysis
(cf.\ \cite{bro:ra489,poriti:eds}, \cite[Ex.\ 5.3.3--24]{knu:acpIII2});
the first version was validated in \cite{kiw:rsq}, and the second one
will be addressed elsewhere.  This paper deals with the third version
of {\sc Select} from \cite{flri:asf}, which operates as follows.  Using
a small random sample, it finds an element $v$ almost sure to be just
above the $k$th if $k<n/2$, or below the $k$th if $k\ge n/2$.
Partitioning $X$ about $v$ leaves $\min\{k,n-k\}+o(n)$ elements on
average for the next recursive call, in which $k$ is near $1$ or $n$
with high probability, so this second call eliminates almost all the
remaining elements.

Apparently this version of {\sc Select} has not been analyzed in the
literature, even in the case of distinct elements.  We first revise it
slightly to simplify our analysis.  Then, without assuming that the
elements are distinct, we show that {\sc Select} needs at most
$n+\min\{k,n-k\}+O(n^{2/3}\ln^{1/3}n)$ comparisons on average, with
$\ln^{1/3}n$ replaced by $\ln^{1/2}n$ for the original samples of
\cite{flri:asf}.  Thus the average cost of {\sc Select} reaches the
lower bounds of $1.5n+o(n)$ for median selection and $1.25n+o(n)$
for selecting an element of random rank.  For the latter task,
{\sc Find} has the bound $2n+o(n)$ when its pivot is set to the
median of a random sample of $s$ elements, with $s\to\infty$,
$s/n\to\infty$ as $n\to\infty$ \cite{maro:oss}; thus {\sc Select}
improves upon {\sc Find} mostly by using $k$, the rank of the element
to be found, for selecting the pivot $v$ in each recursive call.

{\sc Select} can be implemented by using the tripartitioning schemes
of \cite[\S5]{kiw:psq}, which include a modified scheme of
\cite{bemc:esf}; more traditional bipartitioning schemes
\cite[\S2]{kiw:psq} can perform quite poorly in {\sc Select} when
equal elements occur.  We add that the implementation of \cite{flri:asf}
avoids random number generation by assuming that the input file is in
random order, but this results in poor performance on some inputs of
\cite{val:iss}; hence our implementation of {\sc Select} employs
random sampling.

Our computational experience shows that {\sc Select} outperforms even
quite sophisticated implementations of {\sc Find} in both comparison
counts and computing times.  To save space, only selected results are
reported for the version of \cite{val:iss}, but our experience with
other versions on many different inputs was similar.  {\sc Select}
turned out to be more stable than {\sc Find}, having much smaller
variations of solution times and numbers of comparisons.  Quite
suprisingly, contrary to the folklore saying that {\sc Select} is only
asymptotically faster than {\sc Find}, {\sc Select} makes significantly
fewer comparisons even for small inputs
(cf.\ Tab.\ \ref{tab:comp_small}).

To relate our results with those of \cite{kiw:rsq}, let's call
{\sc qSelect} the quintary method of \cite{kiw:rsq} stemming from
\cite[\S2.1]{flri:etb}.  {\sc qSelect} eliminates almost all
elements on its first call by using two pivots, almost sure to be
just below and above the $k$th element, in a quintary partitioning
scheme.  Thus most work occurs on the first call of {\sc qSelect},
which corresponds to the first two calls of {\sc Select}.  Hence
{\sc Select} and {\sc qSelect} share the same efficiency estimates,
and in practice make similarly many comparisons. However, {\sc qSelect}
tends to be slightly faster on median finding: although its quintary
scheme is more complex, most of its work is spent on the first pass
through $X$, whereas {\sc Select} first partitions $X$ and then the
remaining part (about half) of $X$ on its second call to achieve a
similar problem reduction.  On the other hand, {\sc Select} makes
fewer comparisons on small inputs.  Of course, future work should assess
more fully the relative merits of {\sc Select} and {\sc qSelect}.  For
now, the tests reported in \cite{kiw:psq,kiw:rsq} and in \S\ref{s:exp}
suggest that both {\sc Select} and {\sc qSelect} can compete
successfully with refined implementations of {\sc Find}.

The paper is organized as follows.  A general version of {\sc Select} is
introduced in \S\ref{s:alg}, and its basic features are analyzed in
\S\ref{s:sample}.  The average performance of {\sc Select} is studied
in \S\ref{s:average}.  A modification that improves practical
performance is introduced in \S\ref{s:modmed}.
Partitioning schemes are discussed in \S\ref{s:ternpart}.
Finally, our computational results are reported in \S\ref{s:exp}.
%The Appendix contains proofs of certain technical results.
%Finally, we have a conclusion section.

Our notation is fairly standard.
$|A|$ denotes the cardinality of a set $A$.
In a given probability space, $\Prob$ is the probability measure,
$\Exp$ is the mean-value operator and $\Prob[\cdot|{\cal E}]$ is the
probability conditioned on an event ${\cal E}$; the complement of
${\cal E}$ is denoted by ${\cal E}'$.
%
%   *** SECTION 2 ***
\section{The algorithm {\sc Select}}
\label{s:alg}
In this section we describe a general version of {\sc Select} in terms
of two auxiliary functions $s(n)$ and $g(n)$ (the sample size and rank
gap), which will be chosen later.  We omit their arguments in general,
as no confusion can arise.
%
%   *** ALGORITHM 2.1 ***
\begin{algorithm}
\label{alg:sel3}
\rm
\hfil\newline\noindent{\bf {\sc Select}$(X,k)$}
(Selects the $k$th smallest element of $X$, with $1\le k\le n:=|X|$)
\medbreak\noindent{\bf Step 1} ({\em Initiation\/}).
If $n=1$, return $x_1$.
%Choose the sample size $s\in\{1\colon n-1\}$ and gap $g>0$.
Choose the sample size $s\le n-1$ and gap $g>0$.
\medbreak\noindent{\bf Step 2} ({\em Sample selection\/}).
Pick randomly a sample $S:=\{y_1,\ldots,y_s\}$ from $X$.
\medbreak\noindent{\bf Step 3} ({\em Pivot selection\/}).
Let $v$ be the output of {\sc Select}$(S,i_v)$, where
\begin{equation}
i_v:=\left\{\begin{array}{ll}
\rlap{$\min$}\phantom{\max}\left\{\,\lceil ks/n+g\rceil,s\,\right\}&
\mbox{if}\ k<n/2,\\
\max\left\{\,\lceil ks/n-g\rceil,1\,\right\}&
\mbox{if}\ k\ge n/2.
\end{array}\right.
\label{iv}
\end{equation}
\medbreak\noindent{\bf Step 4} ({\em Partitioning\/}).
By comparing each element $x$ of $X\setminus S$ to $v$, partition $X$
into the three sets $L:=\{x\in X:x<v\}$, $E:=\{x\in X:x=v\}$ and
$R:=\{x\in X:v<x\}$.
\medbreak\noindent{\bf Step 5} ({\em Stopping test\/}).
If $|L|<k\le|L\cup E|$, return $v$.
\medbreak\noindent{\bf Step 6} ({\em Reduction\/}).
If $k\le|L|$, set $\hat X:=L$, $\hat n:=|\hat X|$ and $\hat k:=k$;
else set $\hat X:=R$, $\hat n:=|\hat X|$ and $\hat k:=k-|L\cup E|$.
\medbreak\noindent{\bf Step 7} ({\em Recursion\/}).
Return {\sc Select}$(\hat X,\hat k)$.
\end{algorithm}

A few remarks on the algorithm are in order.
%
%   *** REMARKS 2.2 ***
\begin{remarks}
\label{r:sel3}
\rm
(a)
The correctness and finiteness of {\sc Select} stem by induction from
the following observations.  The returns of Steps 1 and 5 deliver the
desired element.  At Step 6, $\hat X$ and $\hat k$ are chosen so that
the $k$th smallest element of $X$ is the $\hat k$th smallest element
of $\hat X$, and $\hat n<n$ (since $v\not\in\hat X$).  Also $|S|<n$ for
the recursive call at Step 3.
\par(b)
When Step 5 returns $v$, {\sc Select} may also return information about
the positions of the elements of $X$ relative to $v$.  For instance, if
$X$ is stored as an array, its $k$ smallest elements may be placed first
via interchanges at Step 4 (cf.\ \S\ref{s:ternpart}).  Hence Step 4 need
only compare $v$ with the elements of $X\setminus S$.
\par(c)
The following elementary property is needed in \S\ref{s:average}.
Let $c_n$ denote the maximum number of comparisons taken by {\sc Select}
on any input of size $n$.  Since Step 3 makes at most $c_s$
comparisons with $s<n$, Step 4 needs at most $n-s$, and Step 7 takes
at most $c_{\hat n}$ with $\hat n<n$, by induction $c_n<\infty$ for
all $n$.
\end{remarks}
%
%   *** SECTION 3 ***
\section{Sampling deviations}
\label{s:sample}
In this section we analyze general features of sampling used by
{\sc Select}.
Our analysis hinges on the following bound on the tail of the
hypergeometric distribution established in \cite{hoe:pis} and
rederived shortly in \cite{chv:thd}.
%
%   *** FACT 3.1 ***
\begin{fact}
\label{f:balls3}
Let\/ $s$ balls be chosen uniformly at random from a set of\/ $n$ balls,
of which\/ $r$ are red, and\/ $r'$ be the random variable representing
the number of red balls drawn.  Let\/ $p:=r/n$.  Then
\begin{equation}
\Prob\left[\,r'\ge ps+g\,\right]\le e^{-2g^2\!/s}\quad\forall g\ge0.
\label{Pexpg}
\end{equation}
\end{fact}

Denote by $x_1^*\le\ldots\le x_n^*$ and $y_1^*\le\ldots\le y_s^*$ the
sorted elements of the input set $X$ and the sample set $S$,
respectively, so that $v=y_{i_v}^*$.  The following result will give
bounds on the position of $v$ in the sorted input sequence.
%
%   *** LEMMA 3.2 ***
\begin{lemma}
\label{l:rankgen}
Suppose\/ $\bar\imath:=\max\{1,\min(\lceil\kappa s\rceil,s)\}$,
$\bar\jmath_l:=\max\{\lceil\kappa n-gn/s\rceil,1\}$, and\/
$\bar\jmath_r:=\min\{\lceil\kappa n+gn/s\rceil,n\}$, where\/
$-g<\kappa s\le s+g$, $1\le s\le n$ and $g\ge0$.  Then\/{\rm:}
\par\indent\rlap{\rm(a)}\hphantom{\rm(a)}
$\Prob[y_{\bar\imath}^*<x_{\bar\jmath_l}^*]\le e^{-2g^2\!/s}$ if\/
$\bar\imath\ge\lceil\kappa s\rceil$.
\par\indent\rlap{\rm(b)}\hphantom{\rm(a)}
$\Prob[x_{\bar\jmath_r}^*<y_{\bar\imath}^*]\le e^{-2g^2\!/s}$ if\/
$\bar\imath\le\lceil\kappa s\rceil$.
\end{lemma}
\begin{proof}
Note that $-g<\kappa s\le s+g$ implies that $\bar\jmath_l\le n$ and
$\bar\jmath_r\ge1$ are well-defined.

(a) If $y_{\bar\imath}^*<x_{\bar\jmath_l}^*$, at least $\bar\imath$
samples satisfy $y_i\le x_r^*$, where
$r:=\max_{x_j^*<x_{\bar\jmath_l}^*}j$.
In the setting of Fact \ref{f:balls3}, we have $r$ red elements
$x_j\le x_r^*$, $ps=rs/n$ and $r'\ge\bar\imath$.  Now,
$1\le r\le\bar\jmath_l-1$ implies
$2\le\bar\jmath_l=\lceil\kappa n-gn/s\rceil<\kappa n-gn/s+1$,
so $-rs/n>-\kappa s+g$.  Hence
$\bar\imath-ps-g>\kappa s-\kappa s+g-g=0$, i.e., $r'>ps+g$.
Thus $\Prob[y_{\bar\imath}^*<x_{\bar\jmath_l}^*]\le e^{-2g^2\!/s}$
by \eqref{Pexpg}.

(b) If $x_{\bar\jmath_r}^*<y_{\bar\imath}^*$, $s-\bar\imath+1$ samples
are at least $x_{\bar\jmath+1}^*$ with
$\bar\jmath:=\max_{x_j^*=x_{\bar\jmath_r}^*}j$.  Thus we have
$r:=n-\bar\jmath$ red elements $x_j\ge x_{\bar\jmath+1}^*$,
$ps=s-\bar\jmath s/n$ and $r'\ge s-\bar\imath+1$.  Since
$\bar\imath<\kappa s+1$ and
$n>\bar\jmath\ge\bar\jmath_r\ge\kappa n+gn/s$,
we get $s-\bar\imath+1-ps-g>\bar\jmath s/n-\kappa s-g\ge\kappa s+g-
\kappa s-g=0$.  Hence $r'>ps+g$ and
$\Prob[x_{\bar\jmath_r}^*<y_{\bar\imath}^*]\le
\Prob[r'\ge ps+g]\le e^{-2g^2\!/s}$ by \eqref{Pexpg}.
\qed
\end{proof}

We now bound the position of $v$ relative to $x_k^*$, $x_{k_l}^*$ and
$x_{k_r}^*$, where
\begin{equation}
k_l:=\max\left\{\,\lceil k-2gn/s\rceil,1\,\right\}
\quad\mbox{and}\quad
k_r:=\min\left\{\,\lceil k+2gn/s\rceil,n\,\right\}.
\label{klkr3}
\end{equation}
%
%   *** COROLLARY 3.3 ***
\begin{corollary}
\label{c:rankdir3}
{\rm(a)}
$\Prob[v<x_k^*]\le e^{-2g^2\!/s}$ if\/ $i_v=\lceil ks/n+g\rceil$
and\/ $k<n/2$.
\par\indent\rlap{\rm(b)}\hphantom{\rm(a)}
$\Prob[x_{k_r}^*<v]\le e^{-2g^2\!/s}$
if\/ $k<n/2$.
\par\indent\rlap{\rm(c)}\hphantom{\rm(a)}
$\Prob[x_k^*<v]\le e^{-2g^2\!/s}$ if\/ $i_v=\lceil ks/n-g\rceil$
and\/ $k\ge n/2$.
\par\indent\rlap{\rm(d)}\hphantom{\rm(a)}
$\Prob[v<x_{k_l}^*]\le e^{-2g^2\!/s}$
if\/ $k\ge n/2$.
\par\indent\rlap{\rm(e)}\hphantom{\rm(a)}
If\/ $k<n/2$, then\/
$i_v\ne\lceil ks/n+g\rceil$ iff\/ $n<k+gn/s${\rm;}
similarly, if\/ $k\ge n/2$, then\/
$i_v\ne\lceil ks/n-g\rceil$ iff\/ $k\le gn/s$.
\end{corollary}
\begin{proof}
Use Lem.\ \ref{l:rankgen} with $\kappa s=ks/n+g$ for (a,b), and
$\kappa s=ks/n-g$ for (c,d).
\qed
\end{proof}
%
%   *** SECTION 4 ***
\section{Average case performance}
\label{s:average}
In this section we analyze the average performance of {\sc Select} for
various sample sizes.
%
%   *** SUBSECTION 4.1 ***
\subsection{Floyd-Rivest's samples}
\label{ss:FRsample}
For positive constants $\alpha$ and $\beta$, consider choosing
$s=s(n)$ and $g=g(n)$ as
\begin{equation}
s:=\min\left\{\lceil\alpha f(n)\rceil,n-1\right\}\ \mbox{and}\
g:=(\beta s\ln n)^{1/2}\ \mbox{with}\ f(n):=n^{2/3}\ln^{1/3}n.
\label{sgf}
\end{equation}
This form of $g$ gives a probability bound
$e^{-2g^2\!/s}=n^{-2\beta}$ for Cor.\ \ref{c:rankdir3}.
To get more feeling, suppose $\alpha=\beta=1$ and $s=f(n)$.
Let $\phi(n):=f(n)/n$.  Then $s/n=g/s=\phi(n)$ and it will be seen
that the recursive call reduces $n$ at least by the factor $4\phi(n)$
on average, i.e., $\phi(n)$ is a contraction factor; note that
$\phi(n)\approx2.4\%$ for $n=10^6$ (cf.\ Tab.\ \ref{tab:fnphin}).
%
%   *** TABLE 4.1 ***
\begin{table}
\caption{Sample size $f(n):=n^{2/3}\ln^{1/3}n$ and relative sample size
$\phi(n):=f(n)/n$.}
\label{tab:fnphin}
\footnotesize
\begin{center}
\begin{tabular}{ccccccccc}
\hline
\vphantom{$1^{2^3}$} % Need more vertical space!
$n$     & $10^3$ & $10^4$ & $10^5$ & $10^6$ & $5\cdot10^6$ & $10^7$
        & $5\cdot10^7$    & $10^8$ \\
\hline
$f(n)$  & 190.449& 972.953& 4864.76& 23995.0&       72287.1& 117248
        & 353885 & 568986 \\
$\phi(n)$
        & .190449& .097295& .048648& .023995&       .014557& .011725
        & .007078& .005690\\
\hline
\end{tabular}
\end{center}
\end{table}
%
%   *** THEOREM 4.1 ***
\begin{theorem}
\label{t:selFR}
Let\/ $C_{nk}$ denote the expected number of comparisons made by
{\sc Select} for $s$ and\/ $g$ chosen as in\/ \eqref{sgf} with\/
$\beta\ge1/6$.  There exists a positive constant\/ $\gamma$ such
that
\begin{equation}
C_{nk}\le n+\min\{\,k,n-k\,\}+\gamma f(n)\quad\forall1\le k\le n.
\label{CnkFR}
\end{equation}
\end{theorem}
\begin{proof}
We need a few preliminary facts.
The function $\phi(t):=f(t)/t=(\ln t/t)^{1/3}$ decreases to $0$ on
$[e,\infty)$, whereas $f(t)$ grows to infinity on $[2,\infty)$.
Let $\delta:=4(\beta/\alpha)^{1/2}$.
Pick $\bar n\ge3$ large enough so that
$e-1\le\alpha f(\bar n)\le\bar n-1$ and $e\le\delta f(\bar n)$.
Let $\bar\alpha:=\alpha+1/f(\bar n)$.
Then, by \eqref{sgf} and the monotonicity of $f$ and $\phi$, we have
for $n\ge\bar n$
\begin{equation}
s\le\bar\alpha f(n)\quad\mbox{and}\quad
f(s)\le\bar\alpha\phi(\bar\alpha f(\bar n))f(n),
\label{sfsFR}
\end{equation}
\begin{equation}
f(\lfloor\delta f(n)\rfloor)\le f(\delta f(n))\le
\delta\phi(\delta f(\bar n))f(n).
\label{flfloordeltaFR}
\end{equation}
For instance, the first inequality of \eqref{sfsFR} yields
$f(s)\le f(\bar\alpha f(n))$, whereas
$$
f(\bar\alpha f(n))=\bar\alpha\phi(\bar\alpha f(n))f(n)\le
\bar\alpha\phi(\bar\alpha f(\bar n))f(n).
$$
Also for $n\ge\bar n$,
we have $s=\lceil\alpha f(n)\rceil=\alpha f(n)+\epsilon$ with
$\epsilon\in[0,1)$ in \eqref{sgf}.  Writing $s=\tilde\alpha f(n)$ with
$\tilde\alpha:=\alpha+\epsilon/f(n)\in[\alpha,\bar\alpha)$, we deduce
from \eqref{sgf} that
\begin{equation}
gn/s=(\beta/\tilde\alpha)^{1/2}f(n)\le(\beta/\alpha)^{1/2}f(n).
\label{gnsboundFR}
\end{equation}
In particular, $4gn/s\le\delta f(n)$, since
$\delta:=4(\beta/\alpha)^{1/2}$.  Next, \eqref{sgf} implies
\begin{equation}
ne^{-2g^2\!/s}\le
n^{1-2\beta}=f(n)n^{1/3-2\beta}\ln^{-1/3}n.
\label{ne2g2sFR}
\end{equation}
Using the monotonicity of $f$ and $\phi$, increase $\bar n$ if necessary
to get for all $n\ge\bar n$
\begin{equation}
2\bar\alpha\phi(\bar\alpha f(\bar n))+
\delta\phi(\delta f(\bar n))+2n^{-2\beta}+
2\max\left\{\,[\delta f(n)]^{2/3-2\beta}n^{-2/3},
n^{-2\beta}\,\right\}\le0.95.
\label{0.95FR}
\end{equation}
By Rem.\ \ref{r:sel3}(c), there is $\gamma$ such that \eqref{CnkFR}
holds for all $n\le\bar n$; increasing $\gamma$ if necessary, and
using the monotonicity of $f$ and the assumption $\beta\ge1/6$,
we have for all $n\ge\bar n$
\begin{equation}
2\bar\alpha+2\delta+5n^{1/3-2\beta}\ln^{-1/3}n+
3\max\left\{\,\delta^{1-2\beta}f(n)^{-2\beta},
n^{1/3-2\beta}\ln^{-1/3}n\,\right\}\le0.05\gamma.
\label{0.05FR}
\end{equation}

Let $n'\ge\bar n$.  Assuming \eqref{CnkFR} holds for all $n\le n'$,
for induction let $n=n'+1$.

We need to consider the following two cases in the first call of
{\sc Select}.

{\em Left case\/}: $k<n/2$.
First, suppose the event
${\cal E}_l:=\{x_k^*\le v\le x_{k_r}^*\}$ occurs.  By the rules of
Steps 4--6, we have $\hat X=L$ (from $x_k^*\le v$), $\hat k=k$ and
$\hat n:=|\hat X|\le k_r-1$ (from $v\le x_{k_r}^*$); since
$k_r<k+2gn/s+1$ by \eqref{klkr3}, we get the two (equivalent) bounds
\begin{equation}
\hat n<k+2gn/s\quad\mbox{and}\quad \hat n-\hat k<2gn/s.
\label{hatnleft}
\end{equation}
Note that if $i_v=\lceil ks/n+g\rceil$ then,
by Cor.\ \ref{c:rankdir3}(a,b), the Boole-Benferroni inequality and the
choice \eqref{sgf}, the complement ${\cal E}_l'$ of ${\cal E}_l$ has
$\Prob[{\cal E}_l']\le2e^{-2g^2\!/s}=2n^{-2\beta}$.
Second, if $i_v\ne\lceil ks/n+g\rceil$, then $n<k+gn/s$
(Cor.\ \ref{c:rankdir3}(e)) combined with $k<n/2$ gives $n<2gn/s$;
hence $\hat n-\hat k<\hat n<n<2gn/s$ implies \eqref{hatnleft}.  Since
also ${\cal E}_l$ implies \eqref{hatnleft}, we have
\begin{equation}
\Prob[{\cal A}_l']\le2n^{-2\beta}\quad\mbox{for}\quad
{\cal A}_l:=\left\{\,\hat n-\hat k<2gn/s\,\right\}.
\label{Al}
\end{equation}

{\em Right case\/}: $k\ge n/2$.
First, suppose the event
${\cal E}_r:=\{x_{k_l}^*\le v\le x_k^*\}$ occurs.  By the rules of
Steps 4--6, we have $\hat X=R$ (from $v\le x_k^*$),
$\hat n-\hat k=n-k$ and $\hat n:=|\hat X|\le n-k_l$ (from
$x_{k_l}^*\le v$); since $k_l\ge k-2gn/s$ by \eqref{klkr3}, we get
the two (equivalent) bounds
\begin{equation}
\hat n\le n-k+2gn/s\quad\mbox{and}\quad\hat k\le2gn/s,
\label{hatnright}
\end{equation}
using $\hat n-\hat k=n-k$.
If $i_v=\lceil ks/n-g\rceil$ then, by Cor.\ \ref{c:rankdir3}(c,d),
the complement ${\cal E}_r'$ of ${\cal E}_r$ has
$\Prob[{\cal E}_r']\le2e^{-2g^2\!/s}=2n^{-2\beta}$.
Second, if $i_v\ne\lceil ks/n-g\rceil$, then $k\le gn/s$
(Cor.\ \ref{c:rankdir3}(e)) combined with $k\ge n/2$ gives
$n\le2gn/s$; hence $\hat k\le\hat n<n\le2gn/s$ implies
\eqref{hatnright}.  Thus
\begin{equation}
\Prob[{\cal A}_r']\le2n^{-2\beta}\quad\mbox{for}\quad
{\cal A}_r:=\left\{\,\hat k\le2gn/s\,\right\}.
\label{Ar}
\end{equation}

Since $k<n-k$ if $k<n/2$, $n-k\le k$ if $k\ge n/2$, \eqref{hatnleft} and
\eqref{hatnright} yield
\begin{equation}
\Prob[{\cal B}']\le2n^{-2\beta}\quad\mbox{for}\quad
{\cal B}:=\left\{\,\hat n\le\min\{\,k,n-k\,\}+2gn/s\,\right\}.
\label{B}
\end{equation}
Note that $\min\{k,n-k\}\le\lfloor n/2\rfloor\le n/2$; this relation
will be used implicitly below.

For the recursive call of Step 7, let $\hat s$, $\hat g$ and
$\hat\imath_v$ denote the quantities generated as in \eqref{sgf} and
\eqref{iv} with $n$ and $k$ replaced by $\hat n$ and $\hat k$, let
$\hat v$ be the pivot found at Step 3, and let $\check X$, $\check n$
and $\check k$ correspond to $\hat X$, $\hat n$ and $\hat k$ at Step 7,
so that $\check n:=|\check X|<\hat n$.

The cost of selecting $v$ and $\hat v$ at Step 3 may be estimated as
\begin{equation}
C_{si_v}+C_{\hat s\hat\imath_v}\le
1.5s+\gamma f(s)+1.5\hat s+\gamma f(\hat s)\le 3s+2\gamma f(s),
\label{CsivFR}
\end{equation}
since $f$ is increasing and \eqref{CnkFR} holds for
$\hat s\le s\le n-1=n'$ (cf.\ \eqref{sgf}) from $\hat n<n$.

Let $c:=n-s$ and $\hat c:=\hat n-\hat s$ denote the costs of Step 4
for the two calls.  Since $0\le\hat c<n$ and
$\Exp\hat c=\Exp[\hat c|{\cal B}]\Prob[{\cal B}]+
\Exp[\hat c|{\cal B}']\Prob[{\cal B}']\le
\Exp[\hat c|{\cal B}]+n\Prob[{\cal B}']$, by \eqref{B} we have
\begin{equation}
c+\Exp\hat c\le n-s+\min\{\,k,n-k\,\}+2gn/s+2n^{1-2\beta}.
\label{cEhatc}
\end{equation}

Using \eqref{CnkFR} again with $\check n<n$,
the cost of finishing up at Step 7 is at most
\begin{equation}
\Exp C_{\check n\check k}\le
\Exp\left[\,1.5\check n+\gamma f(\check n)\,\right]=
1.5\Exp \check n+\gamma\Exp f(\check n).
\label{ECcheckn}
\end{equation}
Thus we need suitable bounds for $\Exp\check n$ and $\Exp f(\check n)$,
which may be derived as follows.

To generalize \eqref{B} to the recursive call, consider the events
\begin{equation}
\hat{\cal B}:=\left\{\,\check n\le\min\{\,\hat k,\hat n-\hat k\,\}+
2\hat g\hat n/\hat s\,\right\}
\quad\mbox{and}\quad
{\cal C}:=\left\{\,\check n\le\lfloor\delta f(n)\rfloor\,\right\}.
\label{hatBC}
\end{equation}
By \eqref{Al} and \eqref{Ar}, $\hat{\cal B}\cap{\cal A}_l$ and
$\hat{\cal B}\cap{\cal A}_r$ imply ${\cal C}$, since
$2gn/s+2\hat g\hat n/\hat s\le\delta f(n)$ by \eqref{gnsboundFR} with
$\hat n<n$ and $\delta:=4(\beta/\alpha)^{1/2}$.  For the recursive
call, proceeding as in the derivation of \eqref{B} with $n$ replaced
by $\hat n=i$, $k$ by $\hat k$, etc., shows that, due to random
sampling,
\begin{equation}
\Prob[\hat{\cal B}'|{\cal A}_l,\hat n=i]\le2i^{-2\beta}
\quad\mbox{and}\quad
\Prob[\hat{\cal B}'|{\cal A}_r,\hat n=i]\le2i^{-2\beta}.
\label{PB'AlB'Ar}
\end{equation}

In the left case of $k<n/2$, using $\check n<n$ and
$\Prob[{\cal A}_l']\le2n^{-2\beta}$ (cf.\ \eqref{Al}), we get
$$
\Exp\check n=\Exp[\check n|{\cal A}_l]\Prob[{\cal A}_l]+
\Exp[\check n|{\cal A}_l']\Prob[{\cal A}_l']\le
\Exp[\check n|{\cal A}_l]+n2n^{-2\beta}.
$$
Partitioning ${\cal A}_l$ into the events
${\cal D}_i:={\cal A}_l\cap\{\hat n=i\}$, $i=0\colon n-1$
($\hat n<n$ always), we have
$$
\Exp[\check n|{\cal A}_l]=\sum_{i=0}^{n-1}
\Exp[\check n|{\cal D}_i]\Prob[{\cal D}_i|{\cal A}_l]\le
\max_{i=0\colon n-1}\Exp[\check n|{\cal D}_i],
$$
where $\Exp[\check n|{\cal D}_i]\le\lfloor\delta f(n)\rfloor$ if
$i\le\lfloor\delta f(n)\rfloor+1$, because $\check n<\hat n$ always.
As for the remaining terms,
$\hat{\cal B}\cap{\cal A}_l\subset{\cal C}$ implies
$\Prob[{\cal C}'|{\cal D}_i]\le\Prob[\hat{\cal B}'|{\cal D}_i]\le
2i^{-2\beta}$ by \eqref{PB'AlB'Ar}, where
${\cal C}:=\{\check n\le\lfloor\delta f(n)\rfloor\}$
and $\check n<\hat n=i$ when the event ${\cal D}_i$ occurs, so
$\Exp[\check n|{\cal D}_i]\le\lfloor\delta f(n)\rfloor+i2i^{-2\beta}$.
Hence
$$
\max_{i=0\colon n-1}\Exp[\check n|{\cal D}_i]\le
\lfloor\delta f(n)\rfloor+
\max_{i=\lfloor\delta f(n)\rfloor+2\colon n-1}2i^{1-2\beta},
$$
where the final term is omitted if $\lfloor\delta f(n)\rfloor>n-3$;
otherwise it is at most
$$
2\max\left\{\,(\lfloor\delta f(n)\rfloor+1)^{1-2\beta},
n^{1-2\beta}\,\right\}\le
2\max\left\{\,\delta^{1-2\beta}f(n)^{-2\beta},
n^{1/3-2\beta}\ln^{-1/3}n\,\right\}f(n),
$$
since $\max_{i=\lfloor\delta f(n)\rfloor+1\colon n}2i^{1-2\beta}$
is bounded as above (consider $\beta\ge1/2$, then $\beta<1/2$ and use
$\delta f(n)<\lfloor\delta f(n)\rfloor+1$, the monotonicity of $f$ and
\eqref{ne2g2sFR} for the final inequality).
Collecting the preceding estimates, we obtain
\begin{equation}
\Exp\check n\le\lfloor\delta f(n)\rfloor+2n^{1-2\beta}+
2\max\left\{\,\delta^{1-2\beta}f(n)^{-2\beta},
n^{1/3-2\beta}\ln^{-1/3}n\,\right\}f(n).
\label{Echeckn}
\end{equation}
Similarly, replacing $\check n$ by $f(\check n)$ in our derivations
and using the monotonicity of $f$ yields
\begin{subequations}
\label{Efcheckn}
\begin{equation}
\Exp f(\check n)\le f(\lfloor\delta f(n)\rfloor)+2f(n)n^{-2\beta}+
\max_{i=\lfloor\delta f(n)\rfloor+2\colon n-1}2f(i)i^{-2\beta},
\label{Efcheckn:a}
\end{equation}
where the final term is omitted if $\lfloor\delta f(n)\rfloor>n-3$;
otherwise it is at most
\begin{equation}
2\max\left\{\,
\frac{f(\lfloor\delta f(n)\rfloor+1)}
{(\lfloor\delta f(n)\rfloor+1)^{2\beta}},
\frac{f(n)}{n^{2\beta}}\,\right\}\le
2\max\left\{\,[\delta f(n)]^{2/3-2\beta}n^{-2/3},
n^{-2\beta}\,\right\}f(n).
\label{Efcheckn:b}
\end{equation}
\end{subequations}
To see this, use the monotonicity of $f$ and the fact that for $i\le n$
(cf.\ \eqref{sgf})
$$
f(i)i^{-2\beta}\!/f(n)=i^{2/3-2\beta}n^{-2/3}(\ln i/\ln n)^{1/3}\le
i^{2/3-2\beta}n^{-2/3}.
$$

For the right case, replace ${\cal A}_l$ by ${\cal A}_r$ in the
preceding paragraph to get \eqref{Echeckn}--\eqref{Efcheckn}.

Add the costs \eqref{CsivFR}, \eqref{cEhatc} and \eqref{ECcheckn},
using \eqref{Echeckn}--\eqref{Efcheckn}, to get
\begin{eqnarray*}
C_{nk}&\le&3s+2\gamma f(s)+n-s+\min\{\,k,n-k\,\}+2gn/s+2n^{1-2\beta}\\
&&{}+1.5\lfloor\delta f(n)\rfloor+3n^{1-2\beta}+
3\max\left\{\,\delta^{1-2\beta}f(n)^{-2\beta},
n^{1/3-2\beta}\ln^{-1/3}n\,\right\}f(n)\\
&&{}+\gamma f(\lfloor\delta f(n)\rfloor)+
2\gamma f(n)n^{-2\beta}+
2\gamma\max\left\{\,[\delta f(n)]^{2/3-2\beta}n^{-2/3},
n^{-2\beta}\,\right\}f(n).
\end{eqnarray*}
Now, using the bounds \eqref{sfsFR}--\eqref{flfloordeltaFR},
$2gn/s\le\frac12\delta f(n)$ (cf.\ \eqref{gnsboundFR}) and
\eqref{ne2g2sFR} gives
\begin{eqnarray*}
\lefteqn{C_{nk}\le n+\min\{\,k,n-k\,\}}\\
&&{}+\Big[2\bar\alpha+2\delta+5n^{1/3-2\beta}\ln^{-1/3}n+
3\max\left\{\,\delta^{1-2\beta}f(n)^{-2\beta},
n^{1/3-2\beta}\ln^{-1/3}n\,\right\}\Big]f(n)\\
&&{}+\left[2\bar\alpha\phi(\bar\alpha f(\bar n))+
\delta\phi(\delta f(\bar n))+2n^{-2\beta}+
2\max\left\{\,[\delta f(n)]^{2/3-2\beta}n^{-2/3},
n^{-2\beta}\,\right\}\right]\gamma f(n).
\end{eqnarray*}
By \eqref{0.95FR}--\eqref{0.05FR}, the two bracketed terms above are
at most $0.05\gamma f(n)$ and $0.95\gamma f(n)$, respectively; thus
\eqref{CnkFR} holds as required.
\qed
\end{proof}
%
%   *** SUBSECTION 4.2 ***
\subsection{Other sampling strategies}
\label{ss:othersample}
We now indicate briefly how to adapt the proof of Thm \ref{t:selFR}
to several variations on \eqref{sgf}; a choice similar to
\eqref{sgfFRsn2/3} below was used in \cite{flri:asf}.
%
%   *** REMARKS 4.2 ***
\begin{remarks}
\label{r:selFR}
\rm
(a)
Theorem \ref{t:selFR} remains true for $\beta\ge1/6$ and
\eqref{sgf} replaced by
\begin{equation}
s:=\min\left\{\left\lceil\alpha n^{2/3}\right\rceil,n-1\right\},\
g:=(\beta s\ln n)^{1/2}\ \mbox{and}\
f(n):=n^{2/3}\ln^{1/2}n.
\label{sgfFRsn2/3}
\end{equation}
Indeed, using $e^{3/2}-1\le\alpha\bar n^{2/3}\le\bar n-1$,
$e^{3/2}\le\delta f(\bar n)$, $\bar\alpha:=\alpha+\bar n^{-2/3}$
and $s=\tilde\alpha n^{2/3}$ with $\tilde\alpha\in[\alpha,\bar\alpha)$
yields \eqref{sfsFR}--\eqref{gnsboundFR} as before, and $\ln^{-1/2}$
replaces $\ln^{-1/3}$ in \eqref{ne2g2sFR}, \eqref{0.05FR} and
\eqref{Echeckn}.
\par(b)
Theorem \ref{t:selFR} holds for the following modification of
\eqref{sgf} with $\epsilon_l>1$
\begin{equation}
s:=\min\left\{\lceil\alpha f(n)\rceil,n-1\right\}\ \mbox{and}\
g:=(\beta s\ln^{\epsilon_l}n)^{1/2}\ \mbox{with}\
f(n):=n^{2/3}\ln^{\epsilon_l/3}n.
\label{sgfFRlneps}
\end{equation}
First, using $e^{\epsilon_l}-1\le\alpha f(\bar n)\le\bar n-1$ and
$e^{\epsilon_l}\le\delta f(\bar n)$ gives
\eqref{sfsFR}--\eqref{gnsboundFR} as before.  Next, fix
$\tilde\beta\ge1/6$.  Let $\beta_n:=\beta\ln^{\epsilon_l-1}n$.
Increase $\bar n$ if necessary so that $\beta_i\ge\tilde\beta$ for
all $i\ge\min\{\bar n,\lceil\delta f(\bar n)\rceil\}$; then
replace $\beta$ by $\tilde\beta$ and $\ln^{-1/3}$ by
$\ln^{-\epsilon_l/3}$ in \eqref{ne2g2sFR} and below.
\par(c)
Several other replacements for \eqref{sgf} may be analyzed as in
\cite[\S\S4.1--4.2]{kiw:rsq}.
\par(d)
None of these choices gives $f(n)$ better than that in \eqref{sgf} for
the bound \eqref{CnkFR}.
\end{remarks}

We now comment briefly on the possible use of sampling with
replacement.
%
%   *** REMARKS 4.3 ***
\begin{remarks}
\label{r:binsample}
\rm
(a)
Suppose Step 2 of {\sc Select} employs sampling with replacement.
Since the tail bound \eqref{Pexpg} remais valid for the binomial
distribution \cite{chv:thd,hoe:pis}, Lemma \ref{l:rankgen} is not
affected.  However, when Step 4 no longer skips comparisons with
the elements of $S$, $-s$ in \eqref{cEhatc} is replaced by $0$; the
resulting change in the bound on $C_{nk}$ only needs replacing
$2\bar\alpha$ in \eqref{0.05FR} by $3\bar\alpha$.  Hence the
preceding results remain valid.
\par(b)
Of course, sampling with replacement needs additional storage for
$S$.  However, the increase in both storage and the number of
comparisons may be tolerated because the sample sizes are relatively
small.
\end{remarks}
%
%   *** SUBSECTION 4.3 ***
\subsection{Handling small subfiles}
\label{ss:subfile}
Since the sampling efficiency decreases when $X$ shrinks, consider the
following modification.  For a fixed cut-off parameter
$n_{\rm cut}\ge1$, let sSelect$(X,k)$ be a ``small-select'' routine that
finds the $k$th smallest element of $X$ in at most $C_{\rm cut}<\infty$
comparisons when $|X|\le n_{\rm cut}$ (even bubble sort will do).  Then
{\sc Select} is modified to start with the following
\medbreak\noindent{\bf Step 0} ({\em Small file case\/}).
If $n:=|X|\le n_{\rm cut}$, return sSelect$(X,k)$.

Our preceding results remain valid for this modification.  In fact it
suffices if $C_{\rm cut}$ bounds the {\em expected\/} number of
comparisons of sSelect$(X,k)$ for $n\le n_{\rm cut}$.  For instance,
\eqref{CnkFR} holds for $n\le n_{\rm cut}$ and $\gamma\ge C_{\rm cut}$,
and by induction as in Rem.\ \ref{r:sel3}(c) we have $C_{nk}<\infty$
for all $n$, which suffices for the proof of Thm \ref{t:selFR}.

Another advantage is that even small $n_{\rm cut}$ ($1000$ say) limits
nicely the stack space for recursion.  Specifically, the tail
recursion of Step 7 is easily eliminated (set $X:=\hat X$, $k:=\hat k$
and go to Step 0), and the calls of Step 3 deal with subsets whose
sizes quickly reach $n_{\rm cut}$.  For example, for the choice of
\eqref{sgf} with $\alpha=1$ and $n_{\rm cut}=600$, at most four
recursive levels occur for $n\le2^{31}\approx2.15\cdot10^9$.
%
%   *** SECTION 5 ***
\section{A modified version}
\label{s:modmed}
We now consider a modification inspired by a remark of
\cite{bro:ra489}.  For $k$ close to $\lceil n/2\rceil$, by symmetry
it is best to choose $v$ as the sample median with
$i_v=\lceil s/2\rceil$, thus attempting to get $v$ close to $x_k^*$
instead of $x_{\lceil k-gn/s\rceil}^*$ or $x_{\lceil k+gn/s\rceil}^*$;
then more elements are eliminated.  Hence we may let
\begin{equation}
i_v:=\left\{\begin{array}{ll}
\lceil ks/n+g\rceil&\mbox{if}\ k<n/2-gn/s,\\
\lceil s/2\rceil&
\mbox{if}\ n/2-gn/s\le k\le n/2+gn/s,\\
\lceil ks/n-g\rceil&\mbox{if}\ k>n/2+gn/s.
\end{array}\right.
\label{iv3}
\end{equation}
Note that \eqref{iv3} coincides with \eqref{iv} in the {\em left\/} case
of $k<n/2-gn/s$ and the {\em right\/} case of $k>n/2+gn/s$, but the
{\em middle\/} case of $n/2-gn/s\le k\le n/2+gn/s$ fixes $i_v$
at the median position $\lceil s/2\rceil$; in fact $i_v$ is the median
of the three values in \eqref{iv3}:
\begin{equation}
i_v:=\max\left\{\,\min\left(\,\lceil ks/n+g\rceil,
\lceil s/2\rceil\,\right),\lceil ks/n-g\rceil\,\right\}.
\label{iv3med}
\end{equation}
Corollary \ref{c:rankdir3} remains valid for the left and right cases.
For the middle case, letting
\begin{equation}
j_l:=\max\left\{\,\lceil n/2-gn/s\rceil,1\,\right\}
\quad\mbox{and}\quad
j_r:=\min\left\{\,\lceil n/2+gn/s\rceil,n\,\right\},
\label{jljr3}
\end{equation}
we obtain from Lemma \ref{l:rankgen} with $\kappa=1/2$ the following
complement of Corollary \ref{c:rankdir3}.
%
%   *** COROLLARY 5.1 ***
\begin{corollary}
\label{c:iv3}
$\Prob[v<x_{j_l}^*]\le e^{-2g^2\!/s}$ and\/
$\Prob[x_{j_r}^*<v]\le e^{-2g^2\!/s}$
if\/ $n/2-gn/s\le k\le n/2+gn/s$.
\end{corollary}
%
%   *** THEOREM 5.2 ***
\begin{theorem}
\label{t:selFRmed}
Theorem\/ {\rm\ref{t:selFR}}
holds for {\sc Select} with Step\/ $3$ using\/ \eqref{iv3}.
\end{theorem}
\begin{proof}
We only indicate how to adapt the proof of Thm \ref{t:selFR} following
\eqref{0.05FR}.  As noted after \eqref{iv3}, the left case now has
$k<n/2-gn/s$ and the right case has $k>n/2+gn/s$, so we only need to
discuss the middle case.

{\em Middle case\/}:
$n/2-gn/s\le k\le n/2+gn/s$.  Suppose
the event ${\cal E}_m:=\{x_{j_l}^*\le v\le x_{j_r}^*\}$ occurs
(note that $\Prob[{\cal E}_m']\le 2e^{-2g^2\!/s}=2n^{-2\beta}$
by Cor.\ \ref{c:iv3}).
If $\hat X=L$ then, by the rules of Steps 4--6, we have $\hat k=k$
and $\hat n\le j_r-1$; since $j_r<n/2+gn/s+1$ by \eqref{jljr3}, we
get $\hat n<n/2+gn/s$.  Hence $k\ge n/2-gn/s$ yields
$\hat n<k+2gn/s$ and $\hat n-\hat k<2gn/s$ as in \eqref{hatnleft}.
Next, if $\hat X=R$ then $\hat n-\hat k=n-k$ and $\hat k:=k-|L\cup E|$,
so $L\cup E=\{x\in X:x\le v\}\ni x_{j_l}^*$ gives $\hat k\le k-j_l$.
Since $k\le n/2+gn/s$ and $j_l\ge n/2-gn/s$ by \eqref{jljr3}, we get
$\hat k\le2gn/s$ and $\hat n\le\hat n-\hat k+2gn/s$ as in
\eqref{hatnright}; further, $\hat n\le n-j_l$ yields
$\hat n\le n/2+gn/s$.  Noticing that $n/2-gn/s\le k\le n/2+gn/s$ implies
$n/2\le\min\{k,n-k\}+gn/s$, we have
$\hat n\le\min\{k,n-k\}+2gn/s$ in both cases.

Thus in the middle case we again have \eqref{B} and hence
\eqref{cEhatc}; further, by \eqref{Al} and \eqref{Ar}, the event
${\cal E}_m\subset{\cal A}_l\cup{\cal A}_r$ is partitioned into
${\cal E}_m\cap{\cal A}_l$ and
${\cal E}_m\cap{\cal A}_l'\cap{\cal A}_r$.

Next, reasoning as before, we see that \eqref{PB'AlB'Ar} and hence
\eqref{Echeckn}--\eqref{Efcheckn} remain valid in the left and right
cases, whereas in the middle case we have
\begin{equation}
\Prob[\hat{\cal B}'|{\cal E}_m,{\cal A}_l,\hat n=i]\le2i^{-2\beta}
\quad\mbox{and}\quad
\Prob[\hat{\cal B}'|{\cal E}_m,{\cal A}_l',{\cal A}_r,\hat n=i]\le
2i^{-2\beta}.
\label{PB'EmAl}
\end{equation}

In the middle case,
$\Exp\check n=\Exp[\check n|{\cal E}_m]\Prob[{\cal E}_m]+
\Exp[\check n|{\cal E}_m']\Prob[{\cal E}_m']$ is bounded by
$\Exp[\check n|{\cal E}_m]+2n^{1-2\beta}$, since
$\Prob[{\cal E}_m']\le2n^{-2\beta}$ and $\check n<n$ always.  Next,
partitioning ${\cal E}_m$ into ${\cal E}_m\cap{\cal A}_l$ and
${\cal E}_m\cap{\cal A}_l'\cap{\cal A}_r$, we obtain
$\Exp[\check n|{\cal E}_m]\le
\max\{\Exp[\check n|{\cal E}_m,{\cal A}_l],
\Exp[\check n|{\cal E}_m,{\cal A}_l',{\cal A}_r]\}$, where
$\Exp[\check n|{\cal E}_m,{\cal A}_l]$ and
$\Exp[\check n|{\cal E}_m,{\cal A}_l',{\cal A}_r]$ may be bounded like
$\Exp[\check n|{\cal A}_l]$ and $\Exp[\check n|{\cal A}_r]$ in the left
and right cases to get \eqref{Echeckn}.
Then \eqref{Efcheckn} is obtained similarly, and the conclusion follows
as before.
\qed
\end{proof}
%
%   *** SECTION 6 ***
\section{Ternary partitions}
\label{s:ternpart}
In this section we discuss ways of implementing {\sc Select} when
the input set is given as an array $x[1\colon n]$.  We employ the
following notation.

Each stage works with a segment $x[l\colon r]$ of the input array
$x[1\colon n]$, where $1\le l\le r\le n$ are such that $x_i<x_l$ for
$i=1\colon l-1$, $x_r<x_i$ for $i=r+1\colon n$, and the $k$th smallest
element of $x[1\colon n]$ is the $(k-l+1)$th smallest element of
$x[l\colon r]$.  The task of {\sc Select} is {\em extended\/}: given
$x[l\colon r]$ and $l\le k\le r$,
{\sc Select}$(x,l,r,k,k_-,k_+)$ permutes $x[l\colon r]$ and finds
$l\le k_-\le k\le k_+\le r$
such that $x_i<x_k$ for all $l\le i<k_-$, $x_i=x_k$ for all
$k_-\le i\le k_+$, $x_i>x_k$ for all $k_+<i\le r$.  The initial call
is {\sc Select}$(x,1,n,k,k_-,k_+)$.

A vector swap denoted by $x[a\colon b]\leftrightarrow x[b+1\colon c]$
means that the first $d:=\min(b+1-a,c-b)$ elements of array
$x[a\colon c]$ are exchanged with its last $d$ elements in arbitrary
order if $d>0$; e.g., we may exchange
$x_{a+i}\leftrightarrow x_{c-i}$ for $0\le i<d$, or
$x_{a+i}\leftrightarrow x_{c-d+1+i}$ for $0\le i<d$.
%
%   *** SUBSECTION 6.1 ***
\subsection{Tripartitioning schemes}
\label{ss:tripart}
For a given pivot $v:=x_l$ from the array $x[l\colon r]$, the following
{\em ternary\/} scheme \cite[\S5.1]{kiw:psq} partitions the array into
three blocks, with $x_m<v$ for $l\le m<a$, $x_m=v$ for $a\le m\le b$,
$x_m>v$ for $b<m\le r$.
After comparing the pivot $v$ to $x_r$ to produce the initial setup
\begin{equation}
\begin{tabular}{llrlrlrr}
\hline
\multicolumn{1}{|c|}{$x=v$} &
\multicolumn{2}{|c|}{$x<v$} &
\multicolumn{2}{|c|}{?} &
\multicolumn{2}{|c|}{$x>v$} &
\multicolumn{1}{|c|}{$x=v$} \\
\hline
\vphantom{$1^{{2^3}^4}$} % Need more vertical space!
$l$ & $p$ & $i$ & & & $j$ & $q$ & $r$\\
\end{tabular}
\label{ternini}
\end{equation}
with $i:=l$ and $j:=r$,
we work with the three inner blocks of the array
\begin{equation}
\begin{tabular}{lllrrr}
\hline
\multicolumn{1}{|c|}{$x=v$} &
\multicolumn{1}{|c|}{$x<v$} &
\multicolumn{2}{|c|}{?} &
\multicolumn{1}{|c|}{$x>v$} &
\multicolumn{1}{|c|}{$x=v$} \\
\hline
\vphantom{$1^{{2^3}^4}$} % Need more vertical space!
$l$ & $p$ & $i$ & $j$ & $q$ & $r$\\
\end{tabular}\ ,
\label{ternbeg}
\end{equation}
until the middle part is empty or just contains an element equal to the
pivot
\begin{equation}
\begin{tabular}{llrclrr}
\hline
\multicolumn{1}{|c|}{$x=v$} &
\multicolumn{2}{|c|}{$x<v$} &
\multicolumn{1}{|c|}{$x=v$} &
\multicolumn{2}{|c|}{$x>v$} &
\multicolumn{1}{|c|}{$x=v$} \\
\hline
\vphantom{$1^{{2^3}^4}$} % Need more vertical space!
$l$ & $p$ & $j$ & & $i$ & $q$ & $r$ \\
\end{tabular}
\label{ternmid}
\end{equation}
(i.e., $j=i-1$ or $j=i-2$),
then swap the ends into the middle for the final arrangement
\begin{equation}
\begin{tabular}{llrr}
\hline
\multicolumn{1}{|c|}{$x<v$} &
\multicolumn{2}{|c|}{$x=v$} &
\multicolumn{1}{|c|}{$x>v$} \\
\hline
\vphantom{$1^{{2^3}^4}$} % Need more vertical space!
$l$ & $a$ & $b$ & $r$\\
\end{tabular}\ .
\label{ternend}
\end{equation}
%
%   *** SCHEME A ***
\begin{scheme}[Safeguarded ternary partition]
\label{sts}
\rm
\begin{description}
\itemsep0pt
\item[]
\item[\ref{sts}1.] [Initialize.]
Set $i:=l$, $p:=i+1$, $j:=r$ and $q:=j-1$.
If $v>x_j$, exchange $x_i\leftrightarrow x_j$ and set $p:=i$;
else if $v<x_j$, set $q:=j$.
\item[\ref{sts}2.] [Increase $i$ until $x_i\ge v$.]
Increase $i$ by $1$; then if $x_i<v$, repeat this step.
\item[\ref{sts}3.] [Decrease $j$ until $x_j\le v$.]
Decrease $j$ by $1$; then if $x_j>v$, repeat this step.
\item[\ref{sts}4.] [Exchange.]
(Here $x_j\le v\le x_i$.)
If $i<j$, exchange $x_i\leftrightarrow x_j$; then
if $x_i=v$, exchange $x_i\leftrightarrow x_p$ and increase $p$ by $1$;
if $x_j=v$, exchange $x_j\leftrightarrow x_q$ and decrease $q$ by $1$;
return to \ref{sts}2.
If $i=j$ (so that $x_i=x_j=v$), increase $i$ by $1$ and
decrease $j$ by $1$.
\item[\ref{sts}5.] [Cleanup.]
Set $a:=l+j-p+1$ and $b:=r-q+i-1$.
Exchange $x[l\colon p-1]\leftrightarrow x[p\colon j]$ and
$x[i\colon q]\leftrightarrow x[q+1\colon r]$.
\end{description}
\end{scheme}

Step \ref{sts}1 ensures that $x_l\le v\le x_r$, so steps \ref{sts}2 and
\ref{sts}3 don't need to test whether $i\le j$.  This scheme makes two
extraneous comparisons (only one when $i=j$ at \ref{sts}4).  Spurious
comparisons are avoided in the following modification
\cite[\S5.3]{kiw:psq} of the scheme of \cite{bemc:esf}
(cf.\ \cite[Ex.\ 5.2.2--41]{knu:acpIII2}),
for which $i=j+1$ in \eqref{ternmid}.
%
%   *** SCHEME B ***
\begin{scheme}[Double-index controlled ternary partition]
\label{stind2}
\rm
\begin{description}
\itemsep0pt
\item[]
\item[\ref{stind2}1.] [Initialize.]
Set $i:=p:=l+1$ and $j:=q:=r$.
\item[\ref{stind2}2.] [Increase $i$ until $x_i>v$.]
If $i\le j$ and $x_i<v$, increase $i$ by $1$ and repeat this step.
If $i\le j$ and $x_i=v$, exchange $x_p\leftrightarrow x_i$, increase
$p$ and $i$ by $1$, and repeat this step.
\item[\ref{stind2}3.] [Decrease $j$ until $x_j<v$.]
If $i<j$ and $x_j>v$, decrease $j$ by $1$ and repeat this step.
If $i<j$ and $x_j=v$, exchange $x_j\leftrightarrow x_q$, decrease
$j$ and $q$ by $1$, and repeat this step.
If $i\ge j$, set $j:=i-1$ and go to \ref{stind2}5.
\item[\ref{stind2}4.] [Exchange.]
Exchange $x_i\leftrightarrow x_j$, increase $i$ by $1$,
decrease $j$ by $1$, and return to \ref{stind2}2.
\item[\ref{stind2}5.] [Cleanup.]
Set $a:=l+i-p$ and $b:=r-q+j$.
Swap $x[l\colon p-1]\leftrightarrow x[p\colon j]$ and
$x[i\colon q]\leftrightarrow x[q+1\colon r]$.
\end{description}
\end{scheme}
%
%   *** SUBSECTION 6.2 ***
\subsection{Preparing for ternary partitions}
\label{ss:preptern}
At Step 1, $r-l+1$ replaces $n$ in finding $s$ and $g$.
At Step 2, it is convenient to place the sample in the initial part of
$x[l\colon r]$ by exchanging $x_i\leftrightarrow x_{i+{\rm rand}(r-i)}$
for $l\le i\le r_s:=l+s-1$, where ${\rm rand}(r-i)$ denotes a random
integer, uniformly distributed between $0$ and $r-i$.

Step 3 uses $i:=k-l+1$ and $m:=r-l+1$ instead of $k$ and $n$
to find the pivot position
\begin{equation}
k_v:=\left\{\begin{array}{ll}
\rlap{$\min$}\phantom{\max}
\left\{\,\lceil l-1+is/m+g\rceil,r_s\,\right\}&
\mbox{if}\ i<m/2,\\
\max\left\{\,\lceil l-1+is/m-g\rceil,l\,\right\}&
\mbox{if}\ i\ge m/2,
\end{array}\right.
\label{kv}
\end{equation}
so that the recursive call of {\sc Select}$(x,l,r_s,k_v,k_v^-,k_v^+)$
produces $v:=x_{k_v}$.

After $v$ has been found, our array looks as follows
\begin{equation}
\begin{tabular}{llrrccr}
\hline
\multicolumn{1}{|c|}{$x<v$} &
\multicolumn{2}{|c|}{$x=v$} &
\multicolumn{1}{|c|}{$x>v$} &
\multicolumn{2}{|c|}{?}\\
\hline
\vphantom{$1^{{2^3}^4}$} % Need more vertical space!
$l$ & $k_v^-$ & $k_v^+$ & $r_s$ & & $r$\\
\end{tabular}\ .
\label{partrec}
\end{equation}
Setting $\bar l:=k_v^-$ and $\bar r:=r-r_s+k_v^+$, we swap
$x[k_v^++1\colon r_s]\leftrightarrow x[r_s+1\colon r]$ in
\eqref{partrec} to get
\begin{equation}
\begin{tabular}{llrlrr}
\hline
\multicolumn{1}{|c|}{$x<v$} &
\multicolumn{2}{|c|}{$x=v$} &
\multicolumn{2}{|c|}{?} &
\multicolumn{1}{|c|}{$x>v$} \\
\hline
\vphantom{$1^{{2^3}^4}$} % Need more vertical space!
$l$ & $\bar l$ & $k_v^+$ & & $\bar r$ & $r$\\
\end{tabular}\ .
\label{partini}
\end{equation}
If $k_v^+=r_s$, we use scheme \ref{sts} with $l$ replaced by $k_v^+$
in \ref{sts}1 (cf.\ \eqref{ternini}) and by $\bar l$ in \ref{sts}5
(cf.\ \eqref{ternmid}); for $k_v^+<r_s$, we set
$i:=k_v^+$, $p:=i+1$, $j:=\bar r+1$, $q:=\bar r$, omit \ref{sts}1
and replace $l$, $r$ by $\bar l$, $\bar r$ in \ref{sts}5.
Similarly, for scheme \ref{stind2}, we replace $l$, $r$ by
$k_v^+$, $\bar r$ in \ref{stind2}1, and by $\bar l$, $\bar r$ in
\ref{stind2}5.

After partitioning $l$ and $r$ are updated by setting $l:=b+1$ if
$a\le k$, $r:=a-1$ if $k\le b$.  If $l\ge r$,
{\sc Select} may return $k_-:=k_+:=k$ if $l=r$, $k_-:=r+1$ and
$k_+:=l-1$ if $l>r$.  Otherwise, instead of calling {\sc Select}
recursively, Step 6 may jump back to Step 1, or to Step 0 if sSelect
is used (cf.\ \S\ref{ss:subfile}).

A simple version of sSelect is obtained if Steps 2 and 3 choose $v:=x_k$
when $r-l+1\le n_{\rm cut}$ (this choice of \cite{flri:asf} works well
in practice, but more sophisticated pivots could be tried); then the
ternary partitioning code can be used by sSelect as well.
%
%   *** SECTION 7 ***
\section{Experimental results}
\label{s:exp}
%
%   *** SUBSECTION 7.1 ***
\subsection{Implemented algorithms}
\label{ss:impl}
An implementation of {\sc Select} was programmed in Fortran 77 and
run on a notebook PC (Pentium 4M 2 GHz, 768 MB RAM) under MS
Windows XP.  The input set $X$ was specified as a double precision
array.  For efficiency, the recursion was removed and small arrays with
$n\le n_{\rm cut}$ were handled as if Steps 2 and 3 chose $v:=x_k$;
the resulting version of sSelect (cf.\ \S\S\ref{ss:subfile} and
\ref{ss:preptern}) typically required less than $3.5n$ comparisons.
The choice of \eqref{sgfFRsn2/3} was employed, with the parameters
$\alpha=0.5$, $\beta=0.25$ and $n_{\rm cut}=600$ as proposed in
\cite{flri:asf}; future work should test other sample sizes and
parameters.
%
%   *** SUBSECTION 7.2 ***
\subsection{Testing examples}
\label{ss:examp}
As in \cite{kiw:rsq}, we used minor modifications of the input sequences
of \cite{val:iss}:
\begin{description}
\itemsep0pt
\item[random]
A random permutation of the integers $1$ through $n$.
\item[onezero]
A random permutation of $\lceil n/2\rceil$ ones and $\lfloor n/2\rfloor$
zeros.
\item[sorted]
The integers $1$ through $n$ in increasing order.
\item[rotated]
A sorted sequence rotated left once; i.e., $(2,3,\ldots,n,1)$.
\item[organpipe]
%The integers $1$ through $n/2$ in increasing order, followed by $n/2$
%through $1$ in decreasing order.
The integers $(1,2,\ldots,n/2,n/2,\ldots,2,1)$.
\item[m3killer]
Musser's ``median-of-3 killer'' sequence with $n=4j$ and $k=n/2$:
$$
\left(\begin{array}{ccccccccccccc}
1&  2 & 3&  4 & \ldots&  k-2& k-1& k& k+1& \ldots& 2k-2& 2k-1& 2k\\
1& k+1& 3& k+3& \ldots& 2k-3& k-1& 2&  4 & \ldots& 2k-2& 2k-1& 2k
\end{array}\right).
$$
\item[twofaced]
Obtained by randomly permuting the
elements of an m3killer sequence in positions $4\lfloor\log_2n\rfloor$
through $n/2-1$ and $n/2+4\lfloor\log_2n\rfloor-1$ through $n-2$.
\end{description}
For each input sequence, its (lower) median element was selected
for $k:=\lceil n/2\rceil$.
%
%   *** SUBSECTION 7.3 ***
\subsection{Computational results}
\label{ss:result}
We varied the input size $n$ from $50{,}000$ to $16{,}000{,}000$.  For
the random, onezero and twofaced sequences, for each input size,
20 instances were randomly generated; for the deterministic
sequences, 20 runs were made to measure the solution time.

The performance of {\sc Select} on randomly generated inputs is
summarized in Table \ref{tab:Selrand},
%
%   *** TABLE 7.1 ***
\begin{table}[t!]
\caption{Performance of {\sc Select} on randomly generated inputs.}
\label{tab:Selrand}
\footnotesize
\begin{center}
\begin{tabular}{lrrrrrrrrrrrrr}
\hline
Sequence &\multicolumn{1}{c}{Size}
&\multicolumn{3}{c}{Time $[{\rm msec}]$%
\vphantom{$1^{2^3}$}} % Need more vertical space!
&\multicolumn{3}{c}{Comparisons $[n]$}
&\multicolumn{1}{c}{$\gamma_{\rm avg}$}
&\multicolumn{1}{c}{$L_{\rm avg}$}
&\multicolumn{1}{c}{$P_{\rm avg}$}
&\multicolumn{1}{c}{$N_{\rm avg}$}
&\multicolumn{1}{c}{$p_{\rm avg}$}
&\multicolumn{1}{c}{$s_{\rm avg}$}\\
&\multicolumn{1}{c}{$n$}
&\multicolumn{1}{c}{avg}&\multicolumn{1}{c}{max}&\multicolumn{1}{c}{min}
&\multicolumn{1}{c}{avg}&\multicolumn{1}{c}{max}&\multicolumn{1}{c}{min}
& &\multicolumn{1}{c}{$[n]$}
&\multicolumn{1}{c}{$[\ln n]$}
&\multicolumn{1}{c}{$[\ln n]$} &
&\multicolumn{1}{c}{$[\%n]$}\\
\hline
%dsel20/dsel20x alpha=0.5 beta=0.25 cutoff=600
random     &  50K
&    2&   10&    0& 1.66& 1.77& 1.61& 1.74& 1.65& 0.46& 0.55& 8.33& 2.59\\
           & 100K
&    3&   10&    0& 1.63& 1.71& 1.55& 1.76& 1.63& 0.60& 0.69& 7.58& 2.12\\
           & 500K
&   13&   20&   10& 1.56& 1.61& 1.54& 1.36& 1.56& 0.67& 0.74& 8.05& 1.19\\
           &   1M
&   23&   30&   20& 1.52& 1.58& 1.00& 0.55& 1.52& 0.66& 0.73& 8.32& 0.91\\
           &   2M
&   46&   51&   40& 1.54& 1.56& 1.52& 1.22& 1.54& 0.75& 0.82& 8.38& 0.72\\
           &   4M
&   88&   91&   80& 1.53& 1.55& 1.52& 1.18& 1.53& 0.86& 0.92& 8.22& 0.57\\
           &   8M
&  172&  181&  160& 1.52& 1.53& 1.51& 1.13& 1.52& 0.92& 0.98& 8.54& 0.44\\
           &  16M
&  336&  341&  320& 1.52& 1.53& 1.51& 1.06& 1.52& 0.95& 1.01& 8.41& 0.35\\
onezero    &  50K
&    2&   10&    0& 1.28& 1.51& 1.00& 0.00& 1.28& 0.24& 0.18& 1.26& 1.91\\
           & 100K
&    3&   10&    0& 1.25& 1.51& 1.00& 0.00& 1.25& 0.26& 0.15& 1.20& 1.49\\
           & 500K
&   15&   20&   10& 1.33& 1.50& 1.00& 0.00& 1.33& 0.29& 0.17& 1.34& 0.93\\
           &   1M
&   30&   41&   20& 1.33& 1.50& 1.00& 0.00& 1.33& 0.27& 0.15& 1.20& 0.73\\
           &   2M
&   60&   71&   41& 1.30& 1.50& 1.00& 0.00& 1.30& 0.26& 0.14& 1.29& 0.56\\
           &   4M
&  109&  131&   90& 1.20& 1.50& 1.00& 0.00& 1.20& 0.22& 0.13& 1.18& 0.41\\
           &   8M
&  219&  261&  190& 1.20& 1.50& 1.00& 0.00& 1.20& 0.22& 0.13& 1.31& 0.32\\
           &  16M
&  436&  501&  370& 1.25& 1.50& 1.00& 0.00& 1.25& 0.20& 0.11& 1.21& 0.27\\
twofaced   &  50K
&    1&   10&    0& 1.67& 1.77& 1.59& 1.87& 1.67& 0.47& 0.56& 8.24& 2.63\\
           & 100K
&    3&   11&    0& 1.62& 1.73& 1.56& 1.67& 1.62& 0.60& 0.69& 7.61& 2.11\\
           & 500K
&   12&   20&   10& 1.56& 1.59& 1.53& 1.23& 1.56& 0.63& 0.71& 8.33& 1.18\\
           &   1M
&   24&   31&   20& 1.55& 1.57& 1.53& 1.23& 1.55& 0.69& 0.76& 8.22& 0.92\\
           &   2M
&   45&   51&   40& 1.54& 1.57& 1.52& 1.23& 1.54& 0.78& 0.85& 8.36& 0.73\\
           &   4M
&   88&   91&   80& 1.53& 1.54& 1.52& 1.17& 1.53& 0.88& 0.94& 8.05& 0.57\\
           &   8M
&  170&  180&  160& 1.52& 1.53& 1.51& 1.12& 1.52& 0.90& 0.97& 8.51& 0.44\\
           &  16M
&  332&  341&  320& 1.52& 1.53& 1.51& 1.04& 1.52& 0.96& 1.02& 8.55& 0.35\\
\hline
\end{tabular}
\end{center}
\end{table}
where the average, maximum and minimum solution times are in
milliseconds, and the comparison counts are in multiples of $n$; e.g.,
column six gives $C_{\rm avg}/n$, where $C_{\rm avg}$ is the average
number of comparisons made over all instances.  Thus
$\gamma_{\rm avg}:=(C_{\rm avg}-1.5n)_+/f(n)$ estimates the constant
$\gamma$ in the bound \eqref{CnkFR}; moreover, we have
$C_{\rm avg}\approx L_{\rm avg}$, where $L_{\rm avg}$ is the average
sum of sizes of partitioned arrays.  Further,
$P_{\rm avg}$ is the average number of {\sc Select} partitions, whereas
$N_{\rm avg}$ is the average number of calls to sSelect and
$p_{\rm avg}$ is the average number of sSelect partitions per call;
both $P_{\rm avg}$ and $N_{\rm avg}$ grow slowly with $\ln n$
(linearly on the onezero inputs).
Finally, $s_{\rm avg}$ is the average sum of sample sizes;
$s_{\rm avg}/n^{2/3}$ drops from $0.95$ for $n=50{\rm K}$ to $0.88$ for
$n=16{\rm M}$ on the random and twofaced inputs, and oscillates about
$0.7$ on the onezero inputs, whereas the initial
$s/n^{2/3}\approx\alpha=0.5$.
The results for the random and twofaced sequences are very similar:
the average solution times grow linearly with $n$ (except for small
inputs whose solution times couldn't be measured accurately), and the
differences between maximum and minimum times are quite small (and also
partly due to the operating system).  Except for the smallest inputs,
the maximum and minimum numbers of comparisons are quite close, and
$C_{\rm avg}$ nicely approaches the theoretical lower bound of $1.5n$;
this is reflected in the values of $\gamma_{\rm avg}$.  The results for
the onezero inputs essentially average two cases: the first pass
eliminates either almost all or about half of the elements.

Table \ref{tab:Seldet} exhibits similar features of {\sc Select} on
the deterministic inputs.
%
%   *** TABLE 7.2 ***
\begin{table}[t!]
\caption{Performance of {\sc Select} on deterministic inputs.}
\label{tab:Seldet}
\footnotesize
\begin{center}
\tabcolsep=0.98\tabcolsep
\begin{tabular}{lrrrrrrrrrrrrr}
\hline
Sequence &\multicolumn{1}{c}{Size}
&\multicolumn{3}{c}{Time $[{\rm msec}]$%
\vphantom{$1^{2^3}$}} % Need more vertical space!
&\multicolumn{3}{c}{Comparisons $[n]$}
&\multicolumn{1}{c}{$\gamma_{\rm avg}$}
&\multicolumn{1}{c}{$L_{\rm avg}$}
&\multicolumn{1}{c}{$P_{\rm avg}$}
&\multicolumn{1}{c}{$N_{\rm avg}$}
&\multicolumn{1}{c}{$p_{\rm avg}$}
&\multicolumn{1}{c}{$s_{\rm avg}$}\\
&\multicolumn{1}{c}{$n$}
&\multicolumn{1}{c}{avg}&\multicolumn{1}{c}{max}&\multicolumn{1}{c}{min}
&\multicolumn{1}{c}{avg}&\multicolumn{1}{c}{max}&\multicolumn{1}{c}{min}
& &\multicolumn{1}{c}{$[n]$}
&\multicolumn{1}{c}{$[\ln n]$}
&\multicolumn{1}{c}{$[\ln n]$} &
&\multicolumn{1}{c}{$[\%n]$}\\
\hline
%dsel10o/dsel10ox alpha=0.5 beta=0.25 cutoff=600
sorted     &  50K
&    1&   10&    0& 1.67& 1.76& 1.59& 1.85& 1.66& 0.48& 0.57& 7.24& 2.65\\
           & 100K
&    2&   10&    0& 1.62& 1.69& 1.55& 1.70& 1.62& 0.60& 0.69& 6.76& 2.12\\
           & 500K
&    8&   10&    0& 1.56& 1.62& 1.53& 1.35& 1.56& 0.67& 0.74& 7.52& 1.19\\
           &   1M
&   15&   20&   10& 1.54& 1.58& 1.53& 1.19& 1.54& 0.68& 0.75& 7.87& 0.92\\
           &   2M
&   27&   31&   20& 1.54& 1.56& 1.52& 1.23& 1.54& 0.74& 0.81& 7.61& 0.73\\
           &   4M
&   51&   61&   40& 1.53& 1.55& 1.52& 1.19& 1.53& 0.87& 0.93& 7.34& 0.57\\
           &   8M
&   98&  111&   90& 1.52& 1.53& 1.51& 1.10& 1.52& 0.89& 0.95& 8.03& 0.44\\
           &  16M
&  186&  200&  170& 1.52& 1.52& 1.51& 1.04& 1.52& 0.95& 1.01& 7.99& 0.35\\
rotated    &  50K
&    1&   10&    0& 1.67& 1.78& 1.59& 1.86& 1.66& 0.48& 0.57& 9.45& 2.64\\
           & 100K
&    2&   10&    0& 1.63& 1.73& 1.58& 1.76& 1.63& 0.61& 0.69& 9.12& 2.12\\
           & 500K
&    8&   10&    0& 1.56& 1.62& 1.54& 1.39& 1.56& 0.65& 0.73&10.03& 1.18\\
           &   1M
&   15&   20&   10& 1.55& 1.58& 1.53& 1.29& 1.55& 0.69& 0.76& 9.56& 0.92\\
           &   2M
&   27&   31&   20& 1.54& 1.55& 1.52& 1.19& 1.54& 0.78& 0.84& 8.69& 0.72\\
           &   4M
&   51&   60&   50& 1.53& 1.54& 1.52& 1.18& 1.53& 0.87& 0.94& 8.92& 0.57\\
           &   8M
&   98&  111&   90& 1.52& 1.53& 1.51& 1.12& 1.52& 0.89& 0.96& 9.29& 0.44\\
           &  16M
&  185&  210&  170& 1.52& 1.53& 1.51& 1.04& 1.52& 0.93& 0.99& 8.96& 0.35\\
organpipe  &  50K
&    1&   10&    0& 1.67& 1.78& 1.59& 1.94& 1.67& 0.45& 0.55& 8.21& 2.62\\
           & 100K
&    3&   10&    0& 1.62& 1.69& 1.57& 1.68& 1.62& 0.60& 0.69& 7.61& 2.11\\
           & 500K
&   10&   10&   10& 1.57& 1.60& 1.54& 1.43& 1.56& 0.67& 0.75& 8.18& 1.19\\
           &   1M
&   20&   20&   10& 1.55& 1.58& 1.52& 1.24& 1.55& 0.70& 0.77& 8.21& 0.93\\
           &   2M
&   37&   41&   30& 1.53& 1.55& 1.52& 1.15& 1.53& 0.78& 0.85& 8.48& 0.72\\
           &   4M
&   68&   80&   60& 1.53& 1.54& 1.52& 1.13& 1.53& 0.84& 0.91& 8.21& 0.57\\
           &   8M
&  130&  150&  120& 1.52& 1.54& 1.51& 1.07& 1.52& 0.88& 0.94& 8.64& 0.44\\
           &  16M
&  240&  260&  230& 1.52& 1.53& 1.51& 1.02& 1.52& 0.94& 1.00& 8.44& 0.35\\
m3killer   &  50K
&    1&   10&    0& 1.67& 1.76& 1.60& 1.89& 1.67& 0.47& 0.55& 8.82& 2.62\\
           & 100K
&    4&   10&    0& 1.63& 1.71& 1.57& 1.80& 1.63& 0.60& 0.69& 7.69& 2.13\\
           & 500K
&   11&   20&   10& 1.57& 1.62& 1.53& 1.44& 1.57& 0.66& 0.73& 8.61& 1.19\\
           &   1M
&   20&   20&   20& 1.55& 1.59& 1.52& 1.40& 1.55& 0.72& 0.79& 8.33& 0.93\\
           &   2M
&   38&   41&   30& 1.54& 1.56& 1.52& 1.25& 1.54& 0.78& 0.85& 8.30& 0.73\\
           &   4M
&   73&   81&   70& 1.53& 1.54& 1.52& 1.28& 1.53& 0.87& 0.94& 8.22& 0.57\\
           &   8M
&  137&  150&  130& 1.52& 1.53& 1.51& 1.05& 1.52& 0.91& 0.97& 8.37& 0.44\\
           &  16M
&  248&  260&  230& 1.52& 1.52& 1.51& 0.96& 1.52& 0.92& 0.97& 8.42& 0.35\\
\hline
\end{tabular}
\end{center}
\end{table}
The results for the sorted and rotated sequences are very similar,
whereas the solution times on the organpipe and m3killer sequences
are between those for the sorted and random sequences.

The results of Tabs.\ \ref{tab:Selrand}--\ref{tab:Seldet} were obtained
with scheme \ref{sts} of \S\ref{ss:preptern}; to save space,
Table \ref{tab:SelpartB} gives only selected results for scheme
\ref{stind2},
%
%   *** TABLE 7.3 ***
\begin{table}[t!]
\caption{Performance of {\sc Select} with ternary scheme \ref{stind2}.}
\label{tab:SelpartB}
\footnotesize
\begin{center}
\begin{tabular}{lrrrrrrrrrrrrr}
\hline
Sequence &\multicolumn{1}{c}{Size}
&\multicolumn{3}{c}{Time $[{\rm msec}]$%
\vphantom{$1^{2^3}$}} % Need more vertical space!
&\multicolumn{3}{c}{Comparisons $[n]$}
&\multicolumn{1}{c}{$\gamma_{\rm avg}$}
&\multicolumn{1}{c}{$L_{\rm avg}$}
&\multicolumn{1}{c}{$P_{\rm avg}$}
&\multicolumn{1}{c}{$N_{\rm avg}$}
&\multicolumn{1}{c}{$p_{\rm avg}$}
&\multicolumn{1}{c}{$s_{\rm avg}$}\\
&\multicolumn{1}{c}{$n$}
&\multicolumn{1}{c}{avg}&\multicolumn{1}{c}{max}&\multicolumn{1}{c}{min}
&\multicolumn{1}{c}{avg}&\multicolumn{1}{c}{max}&\multicolumn{1}{c}{min}
& &\multicolumn{1}{c}{$[n]$}
&\multicolumn{1}{c}{$[\ln n]$}
&\multicolumn{1}{c}{$[\ln n]$} &
&\multicolumn{1}{c}{$[\%n]$}\\
\hline
%dsel20b/dsel20bx alpha=0.5 beta=0.25 cutoff=600
random     &  2M
&   43&   51&   40& 1.53& 1.54& 1.52& 1.02& 1.53& 0.76& 0.83& 8.31& 0.72\\
           &   4M
&   93&  101&   90& 1.53& 1.55& 1.52& 1.09& 1.53& 0.85& 0.92& 8.42& 0.57\\
           &   8M
&  177&  190&  170& 1.52& 1.54& 1.51& 1.03& 1.52& 0.87& 0.93& 8.15& 0.44\\
           &  16M
&  343&  350&  340& 1.51& 1.53& 1.51& 0.88& 1.51& 0.91& 0.97& 8.50& 0.35\\
onezero    &  2M
&   82&   91&   70& 1.30& 1.50& 1.00& 0.00& 1.30& 0.26& 0.14& 1.29& 0.56\\
           &   4M
&  149&  180&  130& 1.20& 1.50& 1.00& 0.00& 1.20& 0.22& 0.13& 1.18& 0.41\\
           &   8M
&  304&  351&  270& 1.20& 1.50& 1.00& 0.00& 1.20& 0.22& 0.13& 1.31& 0.32\\
           &  16M
&  621&  711&  531& 1.25& 1.50& 1.00& 0.00& 1.25& 0.20& 0.11& 1.21& 0.27\\
sorted     &  2M
&   23&   30&   20& 1.54& 1.55& 1.52& 1.18& 1.54& 0.78& 0.85& 7.61& 0.72\\
           &   4M
&   43&   50&   40& 1.53& 1.54& 1.51& 1.18& 1.53& 0.86& 0.92& 7.76& 0.57\\
           &   8M
&   82&   90&   80& 1.52& 1.53& 1.51& 1.10& 1.52& 0.89& 0.95& 8.01& 0.44\\
           &  16M
&  156&  160&  150& 1.52& 1.53& 1.51& 1.04& 1.52& 0.97& 1.03& 8.12& 0.35\\
\hline
\end{tabular}
\end{center}
\end{table}
whereas Table \ref{tab:SelpartB}
%
%   *** TABLE 7.4 ***
\begin{table}%[t!]
\caption{Performance of {\sc Select} with the hybrid scheme of
\cite[\S5.6]{kiw:psq}.}
\label{tab:SelpartI}
\footnotesize
\begin{center}
\begin{tabular}{lrrrrrrrrrrrrr}
\hline
Sequence &\multicolumn{1}{c}{Size}
&\multicolumn{3}{c}{Time $[{\rm msec}]$%
\vphantom{$1^{2^3}$}} % Need more vertical space!
&\multicolumn{3}{c}{Comparisons $[n]$}
&\multicolumn{1}{c}{$\gamma_{\rm avg}$}
&\multicolumn{1}{c}{$L_{\rm avg}$}
&\multicolumn{1}{c}{$P_{\rm avg}$}
&\multicolumn{1}{c}{$N_{\rm avg}$}
&\multicolumn{1}{c}{$p_{\rm avg}$}
&\multicolumn{1}{c}{$s_{\rm avg}$}\\
&\multicolumn{1}{c}{$n$}
&\multicolumn{1}{c}{avg}&\multicolumn{1}{c}{max}&\multicolumn{1}{c}{min}
&\multicolumn{1}{c}{avg}&\multicolumn{1}{c}{max}&\multicolumn{1}{c}{min}
& &\multicolumn{1}{c}{$[n]$}
&\multicolumn{1}{c}{$[\ln n]$}
&\multicolumn{1}{c}{$[\ln n]$} &
&\multicolumn{1}{c}{$[\%n]$}\\
\hline
%dsel20d/dsel20dx alpha=0.5 beta=0.25 cutoff=600
random     &  2M
&   44&   50&   40& 1.53& 1.54& 1.52& 1.03& 1.53& 0.76& 0.83& 8.31& 0.72\\
           &   4M
&   86&  100&   80& 1.53& 1.55& 1.52& 1.10& 1.53& 0.85& 0.92& 8.42& 0.57\\
           &   8M
&  163&  171&  160& 1.52& 1.54& 1.51& 1.03& 1.52& 0.87& 0.93& 8.15& 0.44\\
           &  16M
&  317&  321&  310& 1.51& 1.53& 1.51& 0.88& 1.51& 0.91& 0.97& 8.50& 0.35\\
onezero    &  2M
&   74&   80&   70& 1.30& 1.50& 1.00& 0.00& 1.30& 0.26& 0.14& 1.29& 0.56\\
           &   4M
&  141&  151&  130& 1.20& 1.50& 1.00& 0.00& 1.20& 0.22& 0.13& 1.18& 0.41\\
           &   8M
&  285&  301&  270& 1.20& 1.50& 1.00& 0.00& 1.20& 0.22& 0.13& 1.31& 0.32\\
           &  16M
&  578&  621&  541& 1.25& 1.50& 1.00& 0.00& 1.25& 0.20& 0.11& 1.21& 0.27\\
sorted     &  2M
&   23&   30&   20& 1.54& 1.55& 1.52& 1.18& 1.54& 0.78& 0.85& 7.61& 0.72\\
           &   4M
&   42&   50&   40& 1.53& 1.54& 1.51& 1.19& 1.53& 0.86& 0.92& 7.76& 0.57\\
           &   8M
&   80&   80&   80& 1.52& 1.53& 1.51& 1.11& 1.52& 0.89& 0.95& 8.01& 0.44\\
           &  16M
&  153&  170&  150& 1.52& 1.53& 1.51& 1.04& 1.52& 0.97& 1.03& 8.12& 0.35\\
\hline
\end{tabular}
\end{center}
\end{table}
presents results for the hybrid scheme I of \cite[\S5.6]{kiw:psq},
which combines some features of schemes \ref{sts} and \ref{stind2}.
The hybrid scheme is quite competitive, although slower than scheme
\ref{sts} on the onezero inputs.

The preceding results were obtained with the modified choice \eqref{iv3}
of $i_v$.  For brevity, Table \ref{tab:Seliv} gives results for
{\sc Select} with scheme \ref{sts} and the standard choice \eqref{iv}
of $i_v$ on the random inputs only, since these inputs are most
frequently used in theory and practice for evaluating sorting and
selection methods.
%
%   *** TABLE 7.5 ***
\begin{table}%[t!]
\caption{Performance of {\sc Select} with the standard choice of $i_v$.}
\label{tab:Seliv}
\footnotesize
\begin{center}
\begin{tabular}{lrrrrrrrrrrrrr}
\hline
Sequence &\multicolumn{1}{c}{Size}
&\multicolumn{3}{c}{Time $[{\rm msec}]$%
\vphantom{$1^{2^3}$}} % Need more vertical space!
&\multicolumn{3}{c}{Comparisons $[n]$}
&\multicolumn{1}{c}{$\gamma_{\rm avg}$}
&\multicolumn{1}{c}{$L_{\rm avg}$}
&\multicolumn{1}{c}{$P_{\rm avg}$}
&\multicolumn{1}{c}{$N_{\rm avg}$}
&\multicolumn{1}{c}{$p_{\rm avg}$}
&\multicolumn{1}{c}{$s_{\rm avg}$}\\
&\multicolumn{1}{c}{$n$}
&\multicolumn{1}{c}{avg}&\multicolumn{1}{c}{max}&\multicolumn{1}{c}{min}
&\multicolumn{1}{c}{avg}&\multicolumn{1}{c}{max}&\multicolumn{1}{c}{min}
& &\multicolumn{1}{c}{$[n]$}
&\multicolumn{1}{c}{$[\ln n]$}
&\multicolumn{1}{c}{$[\ln n]$} &
&\multicolumn{1}{c}{$[\%n]$}\\
\hline
%dsel20/dsel20x alpha=0.5 beta=0.25 cutoff=600
random     &  50K
&    4&   10&    0& 1.83& 1.97& 1.74& 3.73& 1.83& 0.57& 0.67& 8.49& 2.96\\
           & 100K
&    4&   10&    0& 1.73& 1.83& 1.61& 3.13& 1.73& 0.73& 0.82& 7.80& 2.32\\
           & 500K
&   14&   20&   10& 1.65& 1.69& 1.61& 3.25& 1.65& 0.82& 0.90& 8.40& 1.30\\
           &   1M
&   25&   30&   20& 1.61& 1.65& 1.58& 2.83& 1.60& 0.89& 0.97& 8.28& 0.99\\
           &   2M
&   46&   50&   40& 1.59& 1.61& 1.56& 2.92& 1.59& 0.99& 1.06& 8.01& 0.77\\
           &   4M
&   90&  100&   80& 1.56& 1.58& 1.54& 2.61& 1.56& 1.15& 1.22& 8.34& 0.60\\
           &   8M
&  174&  181&  170& 1.55& 1.57& 1.54& 2.70& 1.55& 1.21& 1.27& 8.09& 0.47\\
           &  16M
&  341&  351&  330& 1.54& 1.56& 1.53& 2.68& 1.54& 1.21& 1.28& 8.33& 0.36\\
\hline
\end{tabular}
\end{center}
\end{table}
The modified choice typically requires fewer comparisons for small
inputs, but its advantages are less pronounced for larger inputs.
A similar behavior was observed for {\sc Select} with scheme
\ref{stind2}. % and for {\sc bSelect}.

For comparison, Table \ref{tab:qSel} extracts from \cite{kiw:rsq}
some results of {\sc qSelect} for the samples \eqref{sgf}.
%
%   *** TABLE 7.6 ***
\begin{table}
\caption{Performance of quintary {\sc qSelect} on random inputs.}
\label{tab:qSel}
\footnotesize
\begin{center}
\begin{tabular}{lrrrrrrrrrrrrr}
\hline
Sequence &\multicolumn{1}{c}{Size}
&\multicolumn{3}{c}{Time $[{\rm msec}]$%
\vphantom{$1^{2^3}$}} % Need more vertical space!
&\multicolumn{3}{c}{Comparisons $[n]$}
&\multicolumn{1}{c}{$\gamma_{\rm avg}$}
&\multicolumn{1}{c}{$L_{\rm avg}$}
&\multicolumn{1}{c}{$P_{\rm avg}$}
&\multicolumn{1}{c}{$N_{\rm avg}$}
&\multicolumn{1}{c}{$p_{\rm avg}$}
&\multicolumn{1}{c}{$s_{\rm avg}$}\\
&\multicolumn{1}{c}{$n$}
&\multicolumn{1}{c}{avg}&\multicolumn{1}{c}{max}&\multicolumn{1}{c}{min}
&\multicolumn{1}{c}{avg}&\multicolumn{1}{c}{max}&\multicolumn{1}{c}{min}
& &\multicolumn{1}{c}{$[n]$}
&\multicolumn{1}{c}{$[\ln n]$}
&\multicolumn{1}{c}{$[\ln n]$} &
&\multicolumn{1}{c}{$[\%n]$}\\
\hline
%dsel10o/dsel10ox alpha=0.5 beta=0.25 cutoff=600
random     &  50K
&    3&   10&    0& 1.81& 1.85& 1.77& 5.23& 1.22& 0.46& 1.01& 7.62& 4.11\\
           & 100K
&    4&   10&    0& 1.72& 1.76& 1.65& 4.50& 1.15& 0.45& 0.99& 8.05& 3.20\\
           & 500K
&   13&   20&   10& 1.62& 1.63& 1.60& 4.14& 1.08& 0.59& 1.27& 7.59& 1.86\\
           &   1M
&   24&   30&   20& 1.59& 1.60& 1.57& 3.93& 1.06& 0.64& 1.35& 8.18& 1.47\\
           &   2M
&   46&   50&   40& 1.57& 1.58& 1.56& 3.73& 1.04& 0.76& 1.59& 7.67& 1.16\\
           &   4M
&   86&   91&   80& 1.56& 1.56& 1.55& 3.61& 1.03& 0.94& 1.94& 7.21& 0.91\\
           &   8M
&  163&  171&  160& 1.54& 1.55& 1.54& 3.45& 1.03& 0.98& 1.99& 7.45& 0.72\\
           &  16M
&  316&  321&  310& 1.53& 1.54& 1.53& 3.44& 1.02& 0.99& 2.02& 7.55& 0.57\\
\hline
\end{tabular}
\end{center}
\end{table}
As noted in \S\ref{s:intro}, {\sc qSelect} is slightly faster than
{\sc Select} on larger inputs because most of its work occurs on the
first partition (cf.\ $L_{\rm avg}$ in Tabs.\ \ref{tab:Selrand} and
\ref{tab:qSel}).  In Table \ref{tab:riSel}
%
%   *** TABLE 7.7 ***
\begin{table}[t!]
\caption{Performance of {\sc riSelect} on random inputs.}
\label{tab:riSel}
\footnotesize
\begin{center}
\begin{tabular}{lrrrrrrrrrr}
\hline
Sequence &\multicolumn{1}{c}{Size}
&\multicolumn{3}{c}{Time $[{\rm msec}]$%
\vphantom{$1^{2^3}$}} % Need more vertical space!
&\multicolumn{3}{c}{Comparisons $[n]$}
&\multicolumn{1}{c}{$L_{\rm avg}$}
&\multicolumn{1}{c}{$P_{\rm avg}$}
&\multicolumn{1}{c}{$N_{\rm rnd}$}\\
&\multicolumn{1}{c}{$n$}
&\multicolumn{1}{c}{avg}&\multicolumn{1}{c}{max}&\multicolumn{1}{c}{min}
&\multicolumn{1}{c}{avg}&\multicolumn{1}{c}{max}&\multicolumn{1}{c}{min}
&\multicolumn{1}{c}{$[\ln n]$}
&\multicolumn{1}{c}{$[n]$}&\\
\hline
%dsel08
random     &  50K
&    2&   10&    0& 3.10& 4.32& 1.88& 3.10& 1.63& 0.45\\
           & 100K
&    4&   10&    0& 2.61& 4.19& 1.77& 2.61& 1.60& 0.20\\
           & 500K
&   17&   20&   10& 2.91& 4.45& 1.69& 2.91& 1.57& 0.25\\
           &   1M
&   33&   41&   20& 2.81& 3.79& 1.84& 2.81& 1.57& 0.40\\
           &   2M
&   62&   90&   40& 2.60& 3.57& 1.83& 2.60& 1.61& 0.35\\
           &   4M
&  135&  191&   90& 2.86& 4.38& 1.83& 2.86& 1.65& 0.55\\
           &   8M
&  249&  321&  190& 2.60& 3.48& 1.80& 2.60& 1.58& 0.40\\
           &  16M
&  553&  762&  331& 2.99& 4.49& 1.73& 2.99& 1.58& 0.40\\
\hline
\end{tabular}
\end{center}
\end{table}
we give corresponding results for {\sc riSelect}, a Fortran version of
the algorithm of \cite{val:iss}.  For these inputs, {\sc riSelect}
behaves like {\sc Find} with median-of-3 pivots (because the
average numbers of randomization steps, $N_{\rm rnd}$, are negligible);
hence the expected value of $C_{\rm avg}$ is of order $2.75n$
\cite{kimapr:ahf}. 

Our final Table \ref{tab:comp_small}
%
%   *** TABLE 7.8 ***
\begin{table}
\caption{Numbers of comparisons per element made on small random
inputs.}
\label{tab:comp_small}
\footnotesize
\begin{center}
\begin{tabular}{lccccccccccc}
\hline
%dsel20x, dsel10x, dsel08x alpha=0.5 beta=0.25 cutoff=600
Size%
\vphantom{$1^{2^3}$} % Need more vertical space!
&
&  1000&  2500&  5000&  7500& 10000& 12500& 15000& 17500& 20000& 25000\\
\hline
&avg
&  2.48&  2.06&  1.93&  1.87&  1.81&  1.79&  1.77&  1.76&  1.74&  1.71\\
{\sc Select}
&max
&  4.25&  3.03&  2.28&  2.22&  2.09&  2.05&  1.95&  1.93&  1.93&  1.93\\
&min
&  1.55&  1.06&  1.03&  1.64&  1.62&  1.61&  1.64&  1.63&  1.59&  1.60\\
\hline
&avg
&  2.86&  2.55&  2.24&  2.16&  2.07&  2.03&  1.98&  1.98&  1.94&  1.90\\
{\sc qSelect}
&max
&  3.97&  3.55&  2.57&  2.38&  2.28&  2.21&  2.16&  2.13&  2.11&  2.31\\
&min
&  2.29&  1.97&  1.98&  1.95&  1.87&  1.86&  1.82&  1.83&  1.82&  1.75\\
\hline
&avg
&  2.72&  2.85&  2.66&  2.71&  2.72&  2.83&  2.78&  2.75&  2.75&  2.84\\
{\sc riSelect}
&max
&  4.40&  4.51&  4.69&  4.43&  4.62&  4.76&  4.64&  4.40&  5.10&  4.77\\
&min
&  1.68&  1.83&  1.75&  1.59&  1.70&  1.77&  1.78&  1.67&  1.90&  1.71\\
\hline
\end{tabular}
\end{center}
\end{table}
shows that {\sc Select} beats its competitors with respect to the
numbers of comparisons made on small random inputs (100 instances for
each input size $n$).

Our computational results, combined with those in
\cite{kiw:psq,kiw:rsq},
suggest that both {\sc Select} and {\sc qSelect} may compete with
{\sc Find} in practice.

%{\bf Acknowledgment}.  I would like to thank the Associate Editor and
%the two anonymous referees for their helpful comments.
{\bf Acknowledgment}.  I would like to thank Olgierd Hryniewicz,
Roger Koenker, Ronald L. Rivest and John D. Valois for useful
discussions.

%\clearpage

%
%   *** REFERENCES ***
\footnotesize
%\bibliography{kckabbr,kalg,kbk,kck,kint,kth}
%\bibliographystyle{kck}
\newcommand{\etalchar}[1]{$^{#1}$}
\newcommand{\noopsort}[1]{} \newcommand{\printfirst}[2]{#1}
  \newcommand{\singleletter}[1]{#1} \newcommand{\switchargs}[2]{#2#1}
\ifx\undefined\bysame
\newcommand{\bysame}{\leavevmode\hbox to3em{\hrulefill}\,}
\fi

\normalsize
%   *** END OF REFERENCES ***
%
\end{document}